\documentclass{emulateapj}

\usepackage{xcolor}
\usepackage{amsmath}

\slugcomment{\textcolor{white}{blank space}}

\shorttitle{Stellar Aureole Measurements}
\shortauthors{DeVore et al.}

\begin{document}


\title{Retrieving Cirrus Microphysical Properties from Stellar Aureoles}


\author{
John G. DeVore\altaffilmark{1},
Joseph A. Kristl\altaffilmark{2},
Saul A. Rappaport\altaffilmark{3} \\
\vspace{5 mm}
\textit{Journal of Geophysical Research: Atmospheres, in press}
}

\altaffiltext{1}{Visidyne, Inc., Santa Barbara, California, USA; devore@visidyne.com}

\altaffiltext{2}{Visidyne, Inc., Burlington, Massachusetts, USA}

\altaffiltext{3}{37-647 M.I.T.\ Department of Physics, 70 Vassar St., Cambridge, MA, 02139}


\begin{abstract}
The aureoles around stars caused by thin cirrus limit nighttime measurement opportunities for ground-based astronomy but can provide information on high-altitude ice crystals for climate research. In this paper we attempt to demonstrate quantitatively how this works. Aureole profiles can be followed out to $\sim$$0.2^\circ$ from stars and $\sim$$0.5^\circ$ from Jupiter. Interpretation of diffracted starlight is similar to that for sunlight, but emphasizes larger particles. Stellar diffraction profiles are very distinctive, typically being approximately flat out to a critical angle followed by gradually steepening power-law falloff with slope less steep than $-3$. Using the relationship between the phase function for diffraction and the average Fourier transform of the projected area of complex ice crystals we show that defining particle size in terms of average projected area normal to the propagation direction of the starlight leads to a simple, analytic approximation representing large-particle diffraction that is nearly independent of crystal habit. A similar analytic approximation for the diffraction aureole allows it to be separated from the point spread function and the sky background. Multiple scattering is deconvolved using the Hankel transform leading to the diffraction phase function. Application of constrained numerical inversion to the phase function then yields a solution for the particle size distribution in the range between $\sim$$50 \, {\rm \mu m}$ and $\sim$$400 \, {\rm \mu m}$. Stellar aureole measurements can provide one of the very few, as well as least expensive, methods for retrieving cirrus microphysical properties from ground-based observations. 
\end{abstract}

\section{Introduction}
\label{sec:Introduction}

\par Accurate retrieval of the properties of particles in the atmosphere, e.g., ice clouds, is of prime importance to understanding their role in climate change and modeling their effects in global simulations (Comstock et al., 2007). Knowledge of the impact of aerosols on climate change has improved so much that in their contribution to the ``Fourth Assessment Report of the Intergovernmental Panel on Climate Change'' Forster et al. (2007) called AERONET (Holben et al. 1998) a ``significant advancement''. However, the impact of cirrus cloud particles is much less certain because they occur high in the atmosphere and are more difficult to monitor.

\par DeVore et al. (2009) described a new Sun and Aureole Measurement (SAM) instrument that measures the radiance profile of the solar disk and surrounding aureole (from $\sim$$0.6^\circ$ to $8^\circ$) in a narrow spectral band centered on a wavelength of 0.67 $\mu$m. Applying the `diffraction approximation' (presented in their paper) to SAM measurements they were able to retrieve size distributions of cloud particles in the range from 5 to 50 $\mu$m. In practice, SAM has been most useful for collecting information on cirrus clouds because of their thinness (i.e., optical depth $\tau \simeq 0.1$ to $3$), which is required for the appearance of aureoles. 

\par In this paper we report what we believe to be a new method for making and interpreting aureole measurements at night. Although the Moon provides an alternative to the Sun, we think that stars provide the best sources and that planets may have some utility as well. The angular size of the Moon is basically the same as that of the Sun and therefore does not allow for retrieving information on large particles that comes from measurements at small angles. Moreover, both the Moon and planets limit where in the sky measurements can be made, whereas suitable stars are distributed across the entire sky. We have used inexpensive technology to measure starlight diffraction in the angular range from $\lesssim$$0.03^\circ$ to $\sim$$0.2^\circ$, which corresponds to the size range from $\sim$$50 \, {\rm \mu m}$ to $\gtrsim$$400 \, {\rm \mu m}$. Since the climate impact of cirrus is sensitive to its microphysical properties through both particle scattering asymmetry and emissivity/absorptivity (Stephens et al. 1990), it is important to be able to measure the size distibutions of large ice crystals such as those we discuss in this work.

\par Our approach is based on aureole measurements with a lens and a medium-quality astronomical CCD camera. This remote sensing method has advantages over in-situ measurements in that (i) it can readily be carried out on virtually any night when thin cirrus clouds are visible, (ii) it is relatively inexpensive to implement, and (iii) the measurements do not disturb the cloud environment. The aureole approach to monitoring cirrus provides one of the few methods available to retrieve cirrus particle microphysical properties from ground-based observations.

\par \S\ref{sec:StellarAureoleMeasurement} describes the various measurements involved in interpreting stellar aureoles, starting with the determination of two important parameters used in aureole profile interpretation, optical depth (\S\ref{sub:OpticalDepth}) and stellar exo-atmospheric irradiance (\S\ref{sub:StellarExoAtmosphericIrradiance}), followed by a discussion of aureole imagery (\S\ref{sub:AureoleProfiles}). Examples of star images illustrate the difference between the weak aureoles associated with small particles, e.g. thin water clouds, and the strong ones attributable to the large ice crystals in cirrus. 

\par \S\ref{sec:AureoleTheoryAndModeling} presents various aspects of the theory and modeling of diffraction by ice crystals used to interpret the measurements. The reader is reminded that despite the wide range of crystal habits that need to be considered, the phase function representing diffraction can be calculated simply using the Fourier transform. Moreover, this solution illuminates a simple wavelength scaling relationship for aureole profiles. Examination of phase functions derived for different habits with the same size measure leads to the selection of a size metric based on average projected area as being particularly useful. Phase functions representing diffraction from distributions of particle sizes are discussed next, followed by discussion of the potentially important role of multiple scattering and an analytic approximation to represent aureole profiles involving multiple scattering. Deconvolution of multiple scattering is presented in the appendix. Next the steps involved in retrieving the particle size distribution from the phase function are presented (\S\ref{sub:PsdRetrieval}), either by fitting the data with an analytic form or using constrained numerical inversion. Finally, the effects of the point spread function are shown to be taken into account using an appropriate analytic approximation (\S\ref{sub:PointSpreadFunction}).

\par \S\ref{sec:AureoleProfileInterpretation} illustrates the application of theory and modeling to interpret aureole profiles using an example. The measured aureole profile is separated into three components so that the diffraction profile can be distinguished from the point spread function and the sky background (\S\ref{sub:DiffractionProfileExtraction}). Next the phase function is deconvoled from the diffraction profile (\S\ref{sub:PhaseFunctionDeconvolution}) and then the particle size distribution (PSD) is determined (\S\ref{sub:PsdSolution}) from it. The section concludes with some more examples of stellar aureoles covering a range of optical depths (\S\ref{sub:MoreExamples}) and a brief look at an example of an aureole around the planet Jupiter. 

\par \S\ref{sec:Summary} provides a short summary of our work. This is followed by an appendix (\S\ref{sec:MultipleScatteringDeconvolution}) discussing the deconvolution of multiple scattering, where it is approximated as a Poisson process (\S\ref{sub:Poisson}). The resulting series is solved for single scattering, double scattering, and then for all orders (\S\ref{sub:SeriesSolution}).

\section{Stellar Aureole Measurement}
\label{sec:StellarAureoleMeasurement}

\par
\begin{figure}[h!]
\noindent\includegraphics[width=20pc]{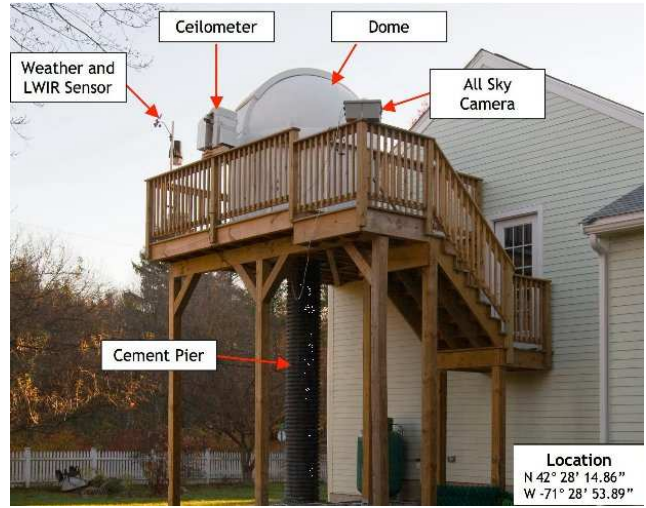}
\caption{West Acton Observatory where the stellar aureole data were collected for Phase I.}
\label{fig:F01_Observatory}
\end{figure}

\begin{figure}[h!]
\noindent\includegraphics[width=20pc]{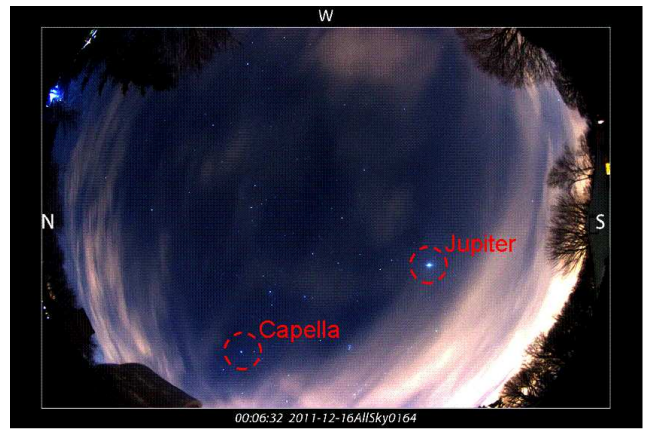}
\caption{Example all-sky image from the evening of 16 Dec 2011 showing the locations of the star Capella and planet Jupiter.}
\label{fig:F02_ExampleAllSkyImage}
\end{figure}

 From our Visidyne measurement site in West Acton, MA (Fig.\,\ref{fig:F01_Observatory}), we collected an extensive stellar aureole dataset comprising approximately 42 hours of imagery from 17 nights between 19 Jun 2011 and 26 Feb 2012 in a narrow 5 nm band around 672 nm. We used multiple exposures to extend the dynamic range. Targets ranged from bright planets such as Jupiter down to magnitude 3 stars. Sky conditions varied from clear to sufficiently cloudy that stars were practically indistinguishable from their aureoles, analogous to the situation with a ``fuzzy Sun'' (Linskens and Bohren, 1994). The clouds consisted primarily of cirrus although water clouds were present at times. The two types of clouds are readily distinguishable from their visual appearance on all-sky imagery as well as from the strength of their aureoles. As with the SAM instrument for solar aureoles, two cameras were used, one to measure particulate optical depth while the second measures the aureole radiance profile at the same time. Supporting data include all-sky imagery, laser cloud altimetry, and both local and synoptic weather data. Fig.\,~\ref{fig:F02_ExampleAllSkyImage} shows a typical image from the all-sky camera with targets identified through the cirrus cloud. 

\subsection{Optical Depth}
\label{sub:OpticalDepth}

\par The central star pixels in the stellar aureole images (discussed in \S\ref{sub:AureoleProfiles}) were saturated in order to set the aureole camera exposure times to yield the best signal for the relatively faint aureole. To measure optical depth at the aureole star, a second, co-aligned camera collected short-exposure, unsaturated images of the aureole star. A $620-680 \, {\rm nm}$ filter was used to allow short exposures (0.05 to 1 second), even during high optical depth conditions. The line-of-sight transmittance was measured by referencing the measured flux to a clear-sky measurement of the same star. The data were scaled by the frame exposure time, and then the optical depth along the line of sight, $\tau_{\rm los}$, was calculated directly as follows:
%
\begin{equation}
\tau_{\rm los} ~= ~- \ln{ ( S_{\rm mea} / S_0 ) }
\label{eqn:OpticalThickness01}
\end{equation}

\noindent where $S_{\rm mea}$ is the measured irradiance and $S_0$ (\S\ref{sub:StellarExoAtmosphericIrradiance}) is a reference measurement on a clear night. Optical depth was measured once per second, and up to 5 depth values were averaged to reduce scatter due to atmospheric scintillation. Determination of optical depth using Sun or star photometry needs to take into account forward scattering. Since the correction is a strong function of the angular width of the source irradiance measurement (e.g., Shiobara and Asano,1994; DeVore et al, 2009), the smaller angular width of stars as compared with the Sun significantly reduces the size of the corrections required. Also, to the extent that the data analysis can distinguish between direct and scattered source radiance, the error attributable to forward scattering can be reduced further. 

\par
\begin{figure}[h!]
\noindent\includegraphics[width=20pc]{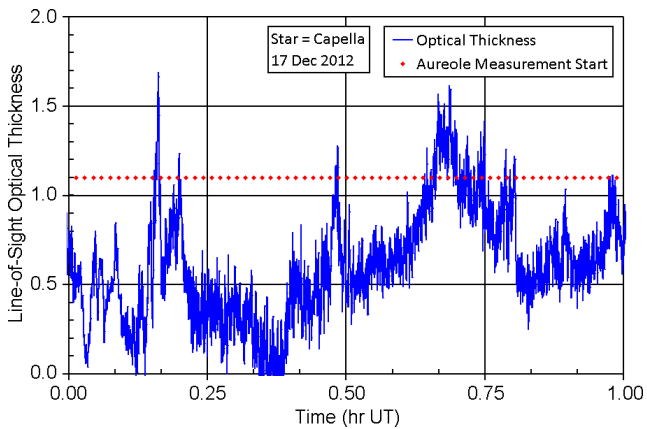}
\caption{Illustrative $\tau_{\rm los}$ measurement data for the star Capella on 17 Dec 2011. The red diamonds indicate the start of 30-second aureole radiance measurements.}
\label{fig:F03_LOS_OpticalDepth}
\end{figure}

 Some typical optical depth data are shown in Fig.\,\ref{fig:F03_LOS_OpticalDepth}. The line-of-sight optical depth is plotted in blue, while the start of each 30-second aureole measurement is indicated with a red diamond. In general, the cirrus clouds are not spatially uniform so that rapid changes in $\tau_{\rm los}$ occur as the clouds drift through the line-of-sight. The high sample rate is used to identify and then discard aureole measurements that contain significant variation (as indicated by the fluctuations in the blue curve) during the longer aureole integration times (as indicated by the separations between the red diamonds). The primary source of error is atmospheric scintillation, which is seen in this type of optical depth measurement even on clear nights. Analysis of clear night data over 30-second time intervals (the duration of aureole image exposures) returns RMS errors of 0.1 to 0.25 in line-of sight optical depth. We recommend use of the conservative error value $\pm 0.125$ for data presented here. Temporal variations caused by cloud motion also limit the total exposure time practical for any single star aureole measurement. 

\subsection{Stellar Exo-Atmospheric Irradiance}
\label{sub:StellarExoAtmosphericIrradiance}

\par As with the determination of $\tau_{\rm los}$, absolute irradiance values are not necessary for specifying $S_0$. It suffices to define irradiance for a specific camera in terms of the number of photons counted divided by the product of the aperture area of the lens $A_{\rm lens} \, ({\rm m^2})$ and the exposure time $t_{\rm exp} \, ({\rm sec)}$, giving units of ${\rm counts / m^2 / sec}$. For radiance one divides the irradiance by the pixel field of view $\Omega_{\rm fov} \, ({\rm sr})$, giving units of ${\rm counts / m^2 / sec / sr}$. 

\par We examined 16 unsaturated images containing Capella from 26 Nov 2011, when the sky was very clear. For each image we determined the maximum, $C_{\rm max}$, and average, $C_{\rm ave}$, number of counts and the corresponding standard deviation, $C_{\rm std}$, in the $21 \times 21$ pixel square surrounding Capella. This size square was large enough so that the minimum number of counts at the edges was down by 2 orders of magnitude from the peak. In order to identify those counts associated with the directly transmitted stellar photons, we separated the individual counts into ``peak'' and ``background'' bins using the following process. First, we selected a somewhat arbitrary threshold value, $C_{\rm thr} = 2 C_{\rm ave}$, to perform an initial separation. Next we calculated the average, $B_{\rm ave}$, and standard deviation, $B_{\rm std}$, of the counts in the background bin. Guided by the three-sigma rule (e.g., Kreyszig, 1979), we reset $C_{\rm thr} = \min ( B_{\rm ave} + 3 B_{\rm std}, C_{\rm max} - 3 B_{\rm std} )$ and redivided the counts into the two bins. Then we calculated the sum of the counts, $C_{\rm peak}$, and the number of pixels, $N_{\rm peak}$, in the peak bin, and recalculated the average number of counts, $B_{\rm avg}$, in the background bin. Then we calculated $S_0 \,{\rm counts / m^2 /sec /sr}$ as:
%
\begin{equation}
S_0 ~= ~\frac{ C_{\rm peak} - N_{\rm peak} B_{\rm ave} }{ A_{\rm lens} \, t_{\rm exp} }
\label{eqn:ExoIrradiance01}
\end{equation}

\par For the 16 images analyzed, the average $S_0 = 2.2 \times 10^8 \,{\rm counts \, m^{-2} s^{-1} sr^{-1}}$ with a standard deviation of $0.1 \times 10^8 \,{\rm counts \, m^{-2} s^{-1} sr^{-1}}$. We selected the maximum of the set, $S_0 = 2.5 \times 10^8 \,{\rm counts \, m^{-2} s^{-1} sr^{-1}}$, to use under the assumption that it represents the clearest sky conditions.

\subsection{Aureole Profiles}
\label{sub:AureoleProfiles}

\par The stellar aureole profiles were collected with a QSI-583wg astronomical camera, interfaced to a Canon 70-200 mm f/\# 2.8 lens set to 200 mm focal length. The camera and lens were mounted to a German equatorial mount that located stars and corrected for sidereal motion. The filter was an Astrodon 672 nm filter with 5 nm bandwidth, chosen to minimize light pollution and provide a narrow band for model calculations. The camera was operated for most collects using $4 \times 4$ pixel binning to enhance the relatively faint aureoles, resulting in single pixel angular sizes of approximately $22^{\prime\prime}$. In later collects the binning was reset to $1 \times 1$, providing higher angular resolution ($5.6^{\prime\prime}$). Each aureole image was a 30 second camera exposure. After the collection, frames were grouped by optical depth (as measured by the optical depth camera) and stacked to improve signal-to-noise. Images that contained significant cloud spatial gradients were discarded prior to stacking.

\par
\begin{figure}[h!]
\noindent\includegraphics[width=20pc]{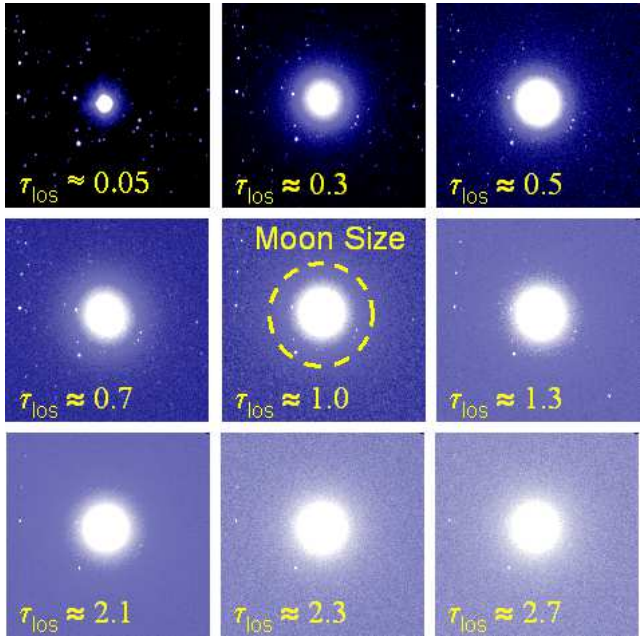}
\caption{Images of aureoles around the star Capella in a narrow 5 nm band around 672 nm during the evening of 16 Dec 2011 annotated with the cirrus optical depth along the line of sight, $\tau_{\rm los}$, as measured by a second camera.}   
\label{fig:F04_CirrusAureoleMontage}
\end{figure}

\begin{figure}[h!]
\noindent\includegraphics[width=20pc]{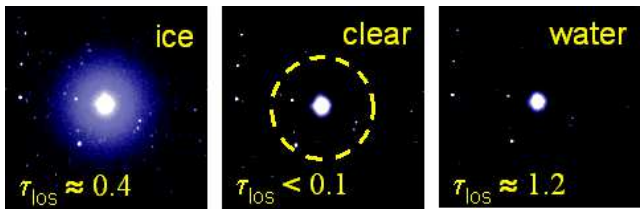}
\caption{Images of the star Capella in a narrow 5 nm band around 672 nm through different cloud conditions during the evening of 11 Jan 2012 annotated with the particulate optical depth along the line of sight, $\tau_{\rm los}$, as measured by a second camera.}   
\label{fig:F05_AureoleExamples}
\end{figure}

 Fig.\,\ref{fig:F04_CirrusAureoleMontage} shows a montage of images of aureoles around the star Capella taken through a thin cirrus layer that varied over the course of the measurements. The images are annotated with $\tau_{\rm los}$ measured by a second camera. Note the expected correlation between the apparent size and intensity of the aureole and $\tau_{\rm los}$. The dashed yellow circle shows the size of the Moon for reference. The overall brightening of the background represents a combination of the ``wings'' of the aureoles, scattering of cityshine from the greater Boston area, and detector noise. Fig.\,\ref{fig:F05_AureoleExamples} compares the images of the star Capella as seen through cirrus, clear sky, and a water cloud on a different night. Note that despite the fact that $\tau_{\rm los}$ in this case is close to the value that should maximize aureole radiance (see below), the water cloud does not produce a discernable aureole. This is due to the fact that small droplets of water diffract light out to angles larger than our field of regard.  (Since the color scalings applied in the two sets of figures differ slightly, one should not place any particular interpretation on the light blue versus black background shading.)

\par
\begin{figure}[h!]
\noindent\includegraphics[width=20pc]{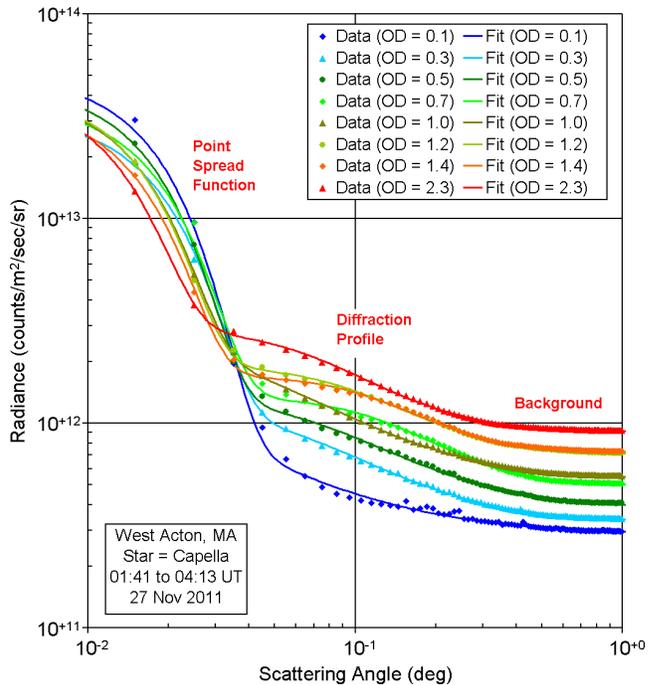}
\caption{Examples of aureole profiles and physical model fits for the star Capella from the night of 27 Nov 2011, annotated by $\tau_{\rm los}$ (``OD'' in the figure legend). }
\label{fig:F06_FirstComponentExamples}
\end{figure}

 Fig.\,\ref{fig:F06_FirstComponentExamples} shows more quantitatively some illustrative aureole profiles about the star Capella as a function of angle. The aureoles span a range of $\tau_{\rm los}$ from 0.1 to 2.3. The colored symbols indicate measured radiances and the solid curves show fits to analytic functions representing the three physical phenomena, the point spread function, stellar diffraction, and sky background, responsible for the shapes of aureole profiles (see \S\,\ref{sec:AureoleProfileInterpretation} for details). The point spread function results from instrumental scattering and atmospheric turbulence and dominates the aureole profiles at small angles;  the resulting radiance is anticorrelated with $\tau_{\rm los}$. The background primarily represents diffusely scattered cityshine and dominates the aureole profiles at large angles; the resulting radiance is positively correlated with $\tau_{\rm los}$. Stellar diffraction gives the aureole its shape at intermediate angles and is the main subject of this paper; the radiance from diffraction peaks for $\tau_{\rm los}$ between about 1 and 2. 

\section{Aureole Theory and Modeling}
\label{sec:AureoleTheoryAndModeling}

\par Various aspects of the theory and modeling of aureoles are covered in this section to provide a basis for their application in the next section. Despite the wide range of crystal habits that need to be considered, the phase function describing diffraction from individual ice crystals can be calculated simply using the Fourier transform (\S\ref{sub:IceCrystalDiffraction}). The solution exhibits a potentially useful wavelength scaling relationship. Airy's analytic solution for a sphere is presented both to provide validation for the implementation of the numerical solution and as a guide for a simpler analytic approximation. Examination of phase functions derived for different habits with the same size measure leads to the selection of a size metric based on the average projected area as being particularly useful. Distributions of particle sizes are discussed next (\S\ref{sub:ParticleSizeDistributions}) considering both  power-law and exponential analytic forms, which leads to a useful analytic approximation for the phase function for a distribution of particles. Next the diffraction radiance profile is considered (\S\ref{sub:RadianceProfiles}) as well as the potentially important role of multiple scattering. Example calculations suggest a useful analytic approximation to represent multiply-scattered diffraction profiles, which can be deconvolved using the equations presented in the appendix. Next the steps involved in retrieving the particle size distribution from the phase function are presented (\S\ref{sub:PsdRetrieval}), either by fitting the data with an analytic form  or using constrained numerical inversion. Finally the effects of the point spread function are taken into account using an appropriate analytic function (\S\ref{sub:PointSpreadFunction}).

\subsection{Diffraction From Individual Ice Crystals}
\label{sub:IceCrystalDiffraction}

\par Diffraction is the dominant mechanism involved in forming aureoles. The angles involved in {\it stellar} aureole measurements tend to be smaller than the size of the Sun, and hence smaller than the angles measured in {\it solar} aureoles. This difference in angular measurement range has important implications for modeling the particles most affecting the aureole shape.  Whereas solar aureole measurements range between $\sim$$0.6^\circ$ and $\sim$$8^\circ$, and correspond to particle dimensions between $\sim$$60 \, {\rm \mu m}$ and $\sim$$5 \, {\rm \mu m}$ according to the diffraction approximation (DeVore et al., 2009), stellar aureole measurements range between $\sim$$0.03^\circ$ and $\sim$$0.2^\circ$ and correspond to particle dimensions an order of magnitude larger. Ice crystals in the sensitive size range for solar aureoles tend to be compact, e.g., droxtals, and hence are well approximated as spheres. However, the ice crystals relevant to stellar aureole modeling require consideration of more complex shapes (e.g., Baum et al. 2005). 

\subsubsection{Diffraction Calculation}
\label{subsub:DiffractionCalculation}

\par
\begin{figure}[h!]
\noindent\includegraphics[width=20pc]{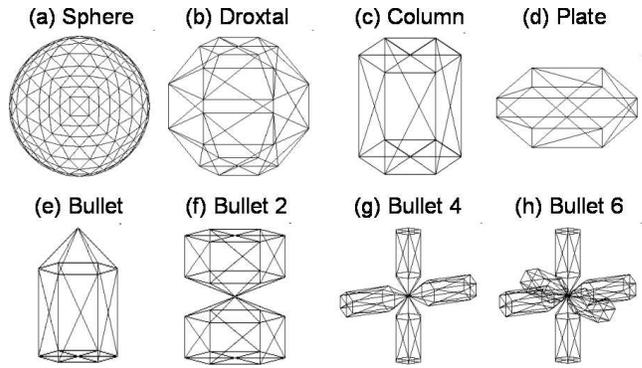}
\caption{Triangular facetized surfaces used to represent 8 different crystal habits. The sphere is included for comparison with the Airy function solution for diffraction from a circular aperture.}
\label{fig:F07_CrystalHabits}
\end{figure}

 Consider diffraction from the 8 crystal habits shown in Fig.\,\ref{fig:F07_CrystalHabits}. Triangular facets are used to describe the outer surfaces of each crystal. Although not representative of a naturally occurring ice crystal we include the spherical shape both as an approximation for small ice crystals and to check our numerical calculations of the diffraction patterns against the well-known solution of Airy (1835) for a sphere. We approximate a sphere (panel a) using a surface with 512 triangular facets, generated by starting with an octahedron and sequentially subdividing each triangle into 3 smaller triangles 3 times, each time moving the vertices to the surface of the circumscribing sphere. Panel (b) shows a droxtal with maximum volume relative to the circumscribing sphere used to represent small, compact ice crystals (Yang et al., 2003). While columns, plates, and bullets are observed, aggregations of bullets are more commonly used to represent the larger ice crystals in cirrus. Panels (c) and (d) show a solid hexagonal column and plate with ratios of the width to length or height based on the scaling relations for crystal habits C1e and P1a provided by Heymsfield and Platt (1984). The bullet and bullet rosettes in panels (e) to (h) use the tip angle and width-to-length scaling relations given by Um and McFarquhar (2007).

\par Bi et al. (2011) discuss several ways of calculating the scattering of light by particles larger than the wavelength $\lambda$ of the incident radiation. Born and Wolf (1959) relate Fraunhofer diffraction to the 2-dimensional Fourier transform of the projection of the particle on the plane normal to the incident direction of the radiation. Although the line integral formulation of Gordon (1975) may be slightly faster, we have found numerical calculation using the fast Fourier transform (Press et al., 1992) adequate and we can calculate the projected area, $\sigma_{\rm pa}$, which we have found useful, in the process. Using the small angle approximation (Lenoble 1985), a phase function $P_{\rm diff}(\theta/\lambda,\phi)$ representing the angular distribution of diffracted radiation can be written as: 
%
\begin{equation}
P_{\rm diff}(\theta/\lambda,\phi) ~= ~\frac{ 2 \pi }{ \sigma_{\rm pa} \, \lambda^2 } ~\vert \mathcal{F} \{ A(x,y) \} \vert^2
\label{eqn:DiffractionPhaseFunction1}
\end{equation}
\noindent where $\theta$ is the scattering angle, $\phi$ is an azimuthal angle, $x$ and $y$ are cartesian coordinates in the projection plane with their origin within the particle projection, $\mathcal{F}$ is the 2-dimensional Fourier transform operator, and $A(x,y)$ is the ``aperture'' function describing the projection of the particle:
%
\begin{equation}
A(x,y) ~= ~1 \, {\rm inside \, the \, projection} \, {\rm and} \, 0 \, {\rm otherwise}
\label{eqn:ApertureFunction1}
\end{equation}

\noindent We represent the aperture function $A(x,y)$ by a two-dimensional, uniformly spaced grid of points. We set the grid dimensions to be 1024 $\times$ 1024 in the examples presented here unless specified otherwise. We set the physical size of the grid in each dimension to the product of a characteristic crystal dimension (e.g., the diameter of an hexagonal plate or the length of a solid column) and the square root of the number of grid points. We project the crystal facets onto a gridded plane oriented normal to the direction of the incident radiation. Each grid point is initialized to zero and set to one if any projected facet covers it. 

\par Unlike an ordinary phase function, $P_{\rm diff}$ is defined only in the forward hemisphere and is meant to represent the diffraction component of the angular distribution of radiation in the near-forward direction. For cases of interest here, i.e., ice crystals large compared with $\lambda$ and $\theta \ll \pi /2$, diffraction dominates the internal or body scattering component, which can be ignored. However, in order to relate $P_{\rm diff}(\theta,\phi)$ correctly to extinction and optical depth $\tau$, both the diffraction and internal components are included in the extinction cross section $\sigma_{\rm ext} = 2 \, \sigma_{\rm pa}$ (from the optical theorem, (e.g., Liou, 2002).


\subsubsection{Wavelength Scaling}
\label{subsub:WavelengthScaling}

\par From Eqn.\,\ref{eqn:DiffractionPhaseFunction1} we see that the amplitude of diffraction phase functions scales inversely with $\lambda^2$ and that the scattering angle $\theta$ scales with $\lambda$. These observations allow for scaling diffraction phase functions in the near forward direction from one wavelength $\lambda_1$ to a second wavelength $\lambda_2$:
%
\begin{eqnarray}
P_{\rm diff}(\theta_2  =  \frac{\lambda_2}{\lambda_1} \theta_1,\phi) ~= \frac{ \lambda_1^2 }{ \lambda_2^2 } ~P_{\rm diff}(\theta_1,\phi) 
\label{eqn:WavelengthScaling1}
\end{eqnarray}
for $ \theta_1 \ll 1$ and $\theta_2 \ll 1$.  The interpretation of Eqn.\,\ref{eqn:WavelengthScaling1} is simple. For a given size crystal as the wavelength increases (decreases) the projected area of the crystal becomes smaller (larger) relative to the wavelength and the scattered radiation spreads to larger (smaller) angles. However, since the total diffracted power remains the same, the radiance at any angle decreases (increases) by the square of the wavelength to conserve energy. Although we shall retain $\lambda$ in a number of equations to emphasize its role, we shall take $\lambda = 0.67 \,{\rm \mu m}$ in the calculations and examples in this work. 

\subsubsection{Airy Solution for a Spherical Particle}
\label{subsub:AirySolutionForASphericalParticle}

\par It is useful to consider the analytic solution for a sphere (Airy 1835). Expressed as a diffraction phase function, Airy's solution for a sphere, $P_{\rm sph}(\theta,D_{\rm sph})$, is given by:
%
\begin{equation}
P_{\rm sph}(\theta,\chi) ~= ~\frac{ \chi^2 }{ 2 } ~\left [ \frac{ 2 J_1( \chi \sin{\theta} ) }{ \chi \sin{\theta} } \right ]^2 ~\simeq ~\frac{ \chi^2 }{ 2 }  ~\left [ \frac{ 2 J_1( \chi \theta ) }{ \chi \theta } ~\right ]^2
\label{eqn:AirySolution01}
\end{equation}
\noindent where $\chi = \pi D_{\rm sph} / \lambda$ is a non-dimensional scaling parameter, $D_{\rm sph}$ is the diameter of the sphere, $J_1$ is the Bessel function of the first kind of order 1, and the small angle approximation ($\sin{\theta} \approx \theta$) has been applied to find the expression on the right.

\par
\begin{figure}[h!]
\noindent\includegraphics[width=20pc]{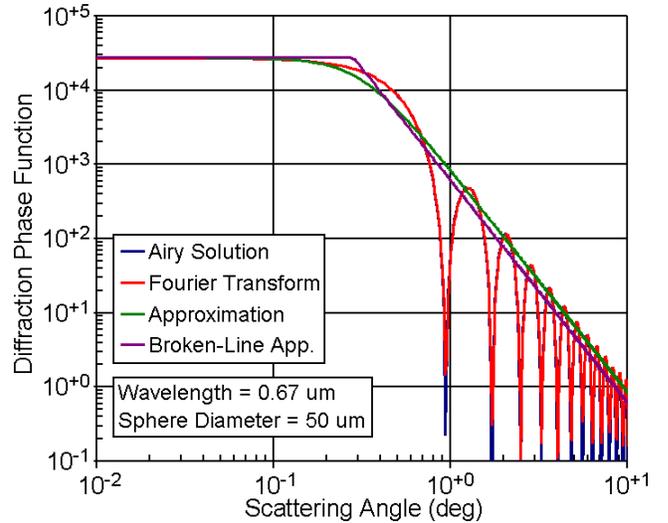}
\caption{Diffraction phase functions calculated for a sphere or facetized sphere of diameter $50 \, {\rm \mu m}$ and $\lambda = 0.67 \, {\rm \mu m}$. Also shown are two analytic approximations.}
\label{fig:F08_PhaseFunctionComparison}
\end{figure}

 Consider the case of light of wavelength $\lambda = 0.67 \,{\rm \mu m}$ incident on a spherical particle [Fig.\,\ref{fig:F07_CrystalHabits} (a)] with diameter $D_{\rm sph} = 50 \, {\rm \mu m}$. The dark blue line in Fig. \ref{fig:F08_PhaseFunctionComparison} shows the Airy solution (Eqn.\,\ref{eqn:AirySolution01}) for the diffraction phase function. We approximated the sphere using a facetted surface as described in \S \ref{subsub:DiffractionCalculation}. However, in this case we subdivided the initial octahedon surface 4 times so that the resulting spherical surface had 2048 facets. Each facet was projected onto a plane of $2048 \times 2048$ points spaced $1.105 \,{\rm \mu m}$ apart. The 1,596 points determined to be inside of projected facets were used to generate an aperture function. They represent a projected area of $1948 \,{\rm \mu m}^2$, which is approximately 1\% less than the projected cross section of a $50 \,{\rm \mu m}$ diameter true sphere. The red line in the figure shows the phase function generated from application of the Fast Fourier transform (Press et al. 1992) to the aperture function using Eqn. \ref{eqn:DiffractionPhaseFunction1}. The two curves are remarkably close but can be distinguished because the minima of the Airy solution are somewhat deeper than those of the numerical one. Although we expect that the limited domain of the numerical Fourier transform is responsible for these differences, they are small compared with the approximations introduced in the following sections. 

\subsubsection{Analytic Approximation}
\label{subsub:AnalyticApproximation}

\par It will prove useful to introduce a simple, analytic approximation, $P_{\rm apx}(\theta,D_{\rm sph})$, to the Airy solution for a sphere in the small angle limit by considering its asymptotic behavior. For small angles, $P_{\rm sph}(\theta,\chi) \rightarrow \chi^2 / 2$, and for large angles, $P_{\rm sph}(\theta,\chi) \propto (\chi \theta)^{-3}$. A simple analytic function $P_{\rm apx}(\theta,\chi)$ with these asymptotic behaviors is:
%
\begin{equation}
P_{\rm apx}(\theta,\chi) ~= ~\frac{ 1 }{ 2 } ~\frac{ \chi^2 }{ 1 + ( \xi \chi \theta )^3 }
\label{eqn:AnalyticApproximation01}
\end{equation}
\noindent where parameter $\xi$ is found from the normalization of $P_{\rm apx}(\theta,\chi)$ using the small angle approximation and taking the upper limit of the integral to infinity:
%
\begin{equation}
\frac{ 1 }{ 2 } ~\int_0^{\infty} ~\frac{ \chi^2 }{ 1 + \xi^3 \chi^3~\theta^3 } ~\theta \, d \theta \, = \,1
\label{eqn:AnalyticApproximation02}
\end{equation}
%
\begin{equation}
\xi ~= ~\frac{ \pi^{1/2} }{ 3^{3/4} } ~\approx ~0.78
\label{eqn:AnalyticApproximation03}
\end{equation}

\noindent The green line in Fig. \ref{fig:F08_PhaseFunctionComparison} shows this analytic approximation. By design, the approximations's plateau at small angles agrees with the Airy and Fourier transform numerical solutions, while at the larger angles the approximation cuts through their oscillations near the peaks. 


\subsubsection{Area Diameter}
\label{subsub:AreaDiameter}

\par
\begin{figure*}
\begin{center}
\includegraphics[width=39pc]{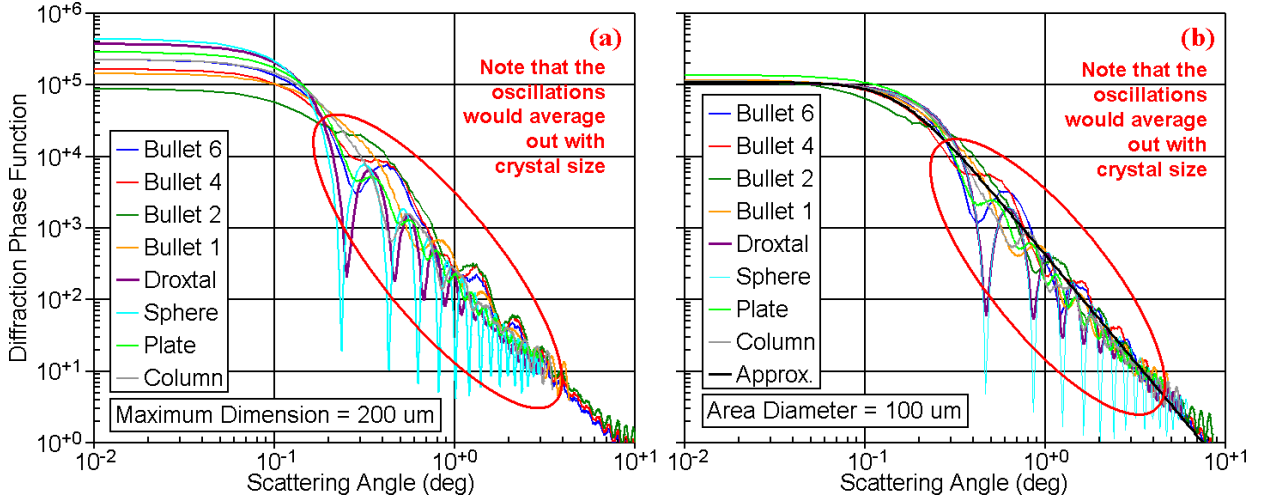}
\caption{Diffraction phase functions calculated for the 8 crystal habits shown in Fig.\,\ref{fig:F07_CrystalHabits} with (a) maximum dimension = $200 \,{\rm \mu m}$ and (b) area diameter = $100 \,{\rm \mu m}$. The latter also shows the analytic approximation (Eqn.\,\ref{eqn:AnalyticApproximation04}) for comparison.}
\label{fig:F09_TwoDimensionsComparison}
\end{center}
\end{figure*}

 Fig.\,\ref{fig:F09_TwoDimensionsComparison} (a) compares calculations of the diffraction phase functions for the 8 crystal habits shown in Fig.\,\ref{fig:F07_CrystalHabits}, all with the {\em same  maximum size}, and averaged over particle orientation. The diffraction phase functions are similar, with plateaus at small scattering angles followed by power-law decreases with slopes of $\sim$$-3$. The oscillations in the power-law regions tend to average out with distributions of particle sizes and/or shapes. However, the overall amplitudes for the same maximum size vary by a factor of $5$ depending upon particle shape. We found that diffraction phase functions for particles with the {\em same volumes} exhibited a factor of 3 variability versus crystal habit. For particles with the {\em same  ratios of their volumes to their average projected areas}, the variability is over an order of magnitude. In other words, none of these measures of particle size proves self-consistent or useful for aureole work. However, the diffraction patterns for particles with the {\em same average projected areas} are very similar as can be seen in Fig.\,\ref{fig:F09_TwoDimensionsComparison} (b), prompting us to define the ``area diameter'' as the diameter of the circle with the same area as that of the average over orientations of the particle projections. The area diameter $D_{\rm a}$ therefore appears to be a useful measure of ice crystal size. This should not be surprising since diffraction by a particle is closely related to its projected area.

\par Given that phase functions are similar for different crystal habits with the same $D_{\rm a}$, including spherical particles, and that the phase functions for the latter can be approximated with a simple analytic function, we are led to replace $D_{\rm sph}$ in the approximation with $D_{\rm a}$. In terms of the scattering angle $\theta$ and diameter $D_{\rm a}$, the analytic approximation in Eqn.\,\ref{eqn:AnalyticApproximation01} becomes:
%
\begin{equation}
P_{\rm apx}(\theta,D_{\rm a}) ~= ~\frac{ 1 }{ 2 } ~\left ( \frac{ \pi ~D_{\rm a} }{ \lambda } \right )^2 ~\frac{ 1 }{ 1 + ~\left ( \frac{ \xi \pi D_{\rm a}  \theta }{ \lambda } \right )^3 }
\label{eqn:AnalyticApproximation04}
\end{equation}
\noindent The black line in Fig.\,\ref{fig:F09_TwoDimensionsComparison} (b) illustrates how well this approximation fits the diffraction calculations. Therefore, for the remainder of this work we adopt the form of Eqn.\,\ref{eqn:AnalyticApproximation04} to represent a ``universal'' phase function for any shape particle with area-diameter $D_{\rm a}$. The extinction cross section $\sigma_{\rm ext}$ is simply twice the projected area:
%
\begin{equation}
\sigma_{\rm ext}(D_{\rm a}) ~= ~2 \frac{ \pi D_{\rm a}^2 }{ 4 } 
\label{eqn:ExtinctionCrossSection01}
\end{equation}

\subsection{Particle Size Distributions}
\label{sub:ParticleSizeDistributions}

\par The simple, analytic approximation $P_{\rm apx}(\theta,D_{\rm a})$ is useful for individual ice crystal habits assuming that they are part of a distribution so that the fine structures of the individual phase functions average out. Next we consider different size distributions of ice crystals and look for ones which are both realistic and have convenient parameterizations. 

\par Using $P_{\rm apx}(\theta,D_{\rm a})$ the total diffraction phase function $P(\theta)$ representing the average over all sizes and habits is given by the following integral over the particle size distribution, PSD:
%
\begin{equation}
P(\theta) = ~\frac{ \int_0^\infty \sigma_{\rm ext}(D_{\rm a}) ~P_{\rm apx}(\theta,D_{\rm a}) ~N(D_{\rm a}) ~d D_{\rm a} }{ \tau_{\rm los} }
\label{eqn:AveragePhaseFunction01}
\end{equation}
\noindent where the extinction (rather than the scattering) cross section has been used (since absorption is negligible at visible wavelengths), $N(D_{\rm a})$ is the differential number density of particles along the path through the atmosphere per unit area diameter, and $\tau_{\rm los}$ is related to $N(D_{\rm a})$ as follows:
%
\begin{equation}
\tau_{\rm los} ~= ~ \int_0^\infty ~\sigma_{\rm ext}(D_{\rm a}) ~N(D_{\rm a}) ~d D_{\rm a}
\label{eqn:OpticalDepthIntegral01}
\end{equation}

\subsubsection{Power-Law Distribution}
\label{subsub:PowerLawDistribution}

\par
\begin{figure*}
\begin{center}
\includegraphics[width=39pc]{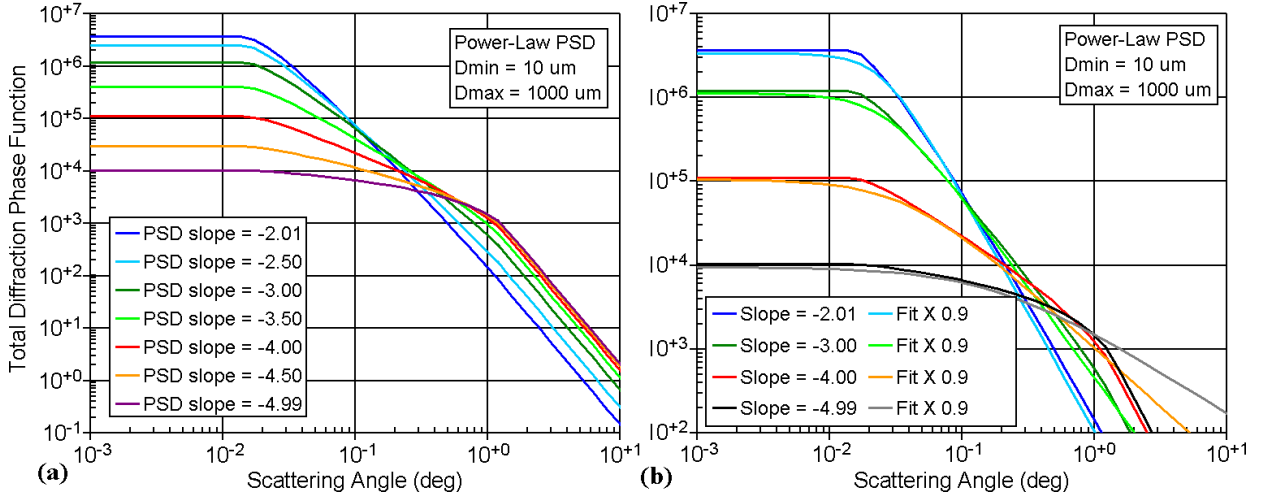}
\caption{ (a) Total diffraction phase functions calculated for a range of $\mu$ from $\approx$$2$ to $\approx$$5$ and (b) comparisons with least-squares fits using the analytic expression in Eqn.\,\ref{eqn:AnalyticApproximation05}.}
\label{fig:F10_PowerLawPhaseFunctions}
\end{center}
\end{figure*}

 Consider a PSD with the power-law form (e.g., Heymsfield and Platt, 1984):   
%
\begin{equation}
N(D_{\rm a}) ~= ~ N_0 \,\, D_{\rm a}^{-\mu} \,\,\,{\rm for \, D_{\rm min} \le D_{\rm a} \le D_{\rm max} }
\label{eqn:PowerLawPSD01}
\end{equation}
\noindent where $N_0$ is a normalization constant, $- \mu$ is the power-law slope, and the PSD is truncated below $D_{\rm a} = D_{\rm min}$ and above $D_{\rm a} = D_{\rm max}$. Substitute Eqn.\,\ref{eqn:PowerLawPSD01} into Eqn.\,\ref{eqn:OpticalDepthIntegral01} and use Eqn.\,\ref{eqn:ExtinctionCrossSection01} to find: 
%
\begin{equation}
N_0 ~= ~\frac{ 2 \, ( 3 - \mu ) \, \tau_{\rm los} }{ \pi \, ( D_{\rm max}^{3-\mu} - D_{\rm min}^{3-\mu} ) }
\label{eqn:PowerLawPSD02}
\end{equation}
\noindent when $\mu \ne 3$. Fig.\,\ref{fig:F10_PowerLawPhaseFunctions} (a) shows numerical integrations of Eqn.\,\ref{eqn:AveragePhaseFunction01} for $2 \lesssim \mu \lesssim 5$ with $D_{\rm a}$ extending from $D_{\rm min} = 10 \,{\rm \mu m}$ to $D_{\rm max} = 1000 \,{\rm \mu m}$ using $P_{\rm apx}(\theta,D_{\rm a})$. The asymptotic behaviors of $P(\theta)$, constant for small $\theta$ and proportional to $\theta^{-3}$ for large $\theta$, are similar to those of the individual analytic phase function $P_{\rm apx}(\theta)$ except that for the steeper PSD slopes there is a transition region with an intermediate power-law slope between $0$ and $-3$. 

\par Although an analytic expression can be found for $P_{\rm apx}(\theta)$, it involves hypergeometric functions and is not very insightful. However, it is illuminating to apply a different approximation to the single particle phase function by replacing $P_{\rm apx}(\theta,D_{\rm a})$ with what we call the broken-line power-law approximation $P_{\rm bp}(\theta,D_{\rm a})$:
%
\begin{equation}
P_{\rm bp}(\theta,D_{\rm a}) ~= ~\frac{ 1 }{ 2 } ~\left ( \frac{ \pi D_{\rm a} }{ \lambda } ~\right )^2 \,\,\, {\rm if \,\, \theta \le \theta_{\rm bp}}
\label{eqn:Broken01}
\end{equation}
\noindent and
%
\begin{equation}
P_{\rm bp}(\theta,D_{\rm a}) ~= ~\frac{ 1 }{ 2 } ~\left ( \frac{ \pi D_{\rm a} }{ \lambda } ~\right )^2 \left ( \frac{ \theta_{\rm bp} }{ \theta } \right )^3
\,\,\, {\rm if \,\, \theta \ge \theta_{\rm bp}}
\label{eqn:Broken02}
\end{equation}
\noindent where $\theta_{\rm bp}$ is determined from the normalization requirement for $P_{\rm bp}(\theta,D_{\rm a})$ to be:
%
\begin{equation}
\theta_{\rm bp} ~= ~\frac{ 2 \lambda }{\sqrt{ 3 } \pi D_{\rm a} }
\label{eqn:Broken03}
\end{equation}
\noindent Fig.\,\ref{fig:F08_PhaseFunctionComparison} compares this approximation (the purple curve) with the continuous version (the green curve) from Eqn.\,\ref{eqn:AnalyticApproximation04}. 

\par With the limits of the integral in Eqn.\,\ref{eqn:AveragePhaseFunction01} again set to $D_{\rm min}$ and $D_{\rm max}$ the analytic result for the total phase function $P_{\rm bp}(\theta)$ naturally divides into three regions depending upon $\theta$ as follows:
%
\begin{eqnarray}
P_{\rm bp}(\theta) ~&=&  ~\frac{ \pi^2 ( 3 - \mu ) ( D_{\rm max}^{5-\mu} - D_{\rm min}^{5-\mu} ) }{ 2 \lambda^2 ( 5 - \mu ) ( D_{\rm max}^{3-\mu} - D_{\rm min}^{3-\mu} ) } \nonumber \\
 && {\rm if \,\, \theta < 2 \lambda / ( \sqrt{ 3 } \pi D_{\rm max} ) }
\label{eqn:Broken04}
\end{eqnarray}
\noindent and
%
\begin{eqnarray}
P_{\rm bp}(\theta) ~&=& ~\frac{ 4 \lambda ( 3 - \mu ) ( D_{\rm max}^{2-\mu} - D_{\rm min}^{2-\mu} ) }{ 3 \sqrt{ 3 } \pi ( 2 - \mu ) ( D_{\rm max}^{3-\mu} - D_{\rm min}^{3-\mu} ) } ~\frac{ 1 }{  \theta^3 } \nonumber \\
 && {\rm if \,\, \theta > 2 \lambda / ( \sqrt{ 3 } \pi D_{\rm min} ) }
\label{eqn:Broken05}
\end{eqnarray}
\noindent and
%
\begin{eqnarray}
P_{\rm bp}(\theta) ~&=&  ~\frac{ \pi^2 ( 3 - \mu ) ( D_{\rm mid}^{5-\mu} - D_{\rm min}^{5-\mu} ) }{ 2 \lambda^2 ( 5 - \mu ) ( D_{\rm max}^{3-\mu} - D_{\rm min}^{3-\mu} ) } \nonumber \\
&+& ~\frac{ 4 \lambda ( 3 - \mu ) ( D_{\rm max}^{2-\mu} - D_{\rm mid}^{2-\mu} ) }{ 3 \sqrt{ 3 } \pi ( 2 - \mu ) ( D_{\rm max}^{3-\mu} - D_{\rm min}^{3-\mu} ) } ~\frac{ 1 }{ \theta^3 } \nonumber \\
 && {\rm otherwise }
\label{eqn:Broken06}
\end{eqnarray}
\noindent where, importantly, $D_{\rm mid} \equiv 2 \lambda / ( \sqrt{ 3 } \pi \theta )$. Substituting this expression into Eqn.\,\ref{eqn:Broken06} gives an equation with the following form for the transition region: 
%
\begin{eqnarray}
P_{\rm bp}(\theta) ~&=& ~c_1 \theta^{\mu-5} ~- ~c_2 \theta^{-3} ~+ ~c_3 \nonumber \\
 && {\rm if \,\, \theta > 2 \lambda / ( \sqrt{ 3 } \pi D_{\rm max} ) } \nonumber \\
 && {\rm and \,\, \theta < 2 \lambda / ( \sqrt{ 3 } \pi D_{\rm min} ) } 
\label{eqn:Broken07}
\end{eqnarray}
\noindent where $c_1$, $c_2$, and $c_3$ are constants that depend upon $D_{\rm min}$, $D_{\rm max}$, and $\mu$. Eqns. \ref{eqn:Broken04} to \ref{eqn:Broken07} show a total phase function which is constant for small $\theta$ and proportional to $\theta^{-3}$ for large $\theta$. If the intermediate transition region of the phase function is modeled with a power-law form, i.e., $P(\theta) \propto \theta^{-\nu}$, then $\nu \approx 5 -\mu$. This result is consistent with the relationship between the phase function (aureole) and PSD power-law slopes found using the diffraction approximation (DeVore et al, 2009).

\subsubsection{Exponential Distribution}
\label{subsub:ExponentialDistribution}

\par For later use we note that an exponential form is also sometimes used to model ice crystal PSDs (e.g., Field and Heymsfield, 2003):
 
%
\begin{equation}
N(D_{\rm a}) ~= ~ N_0 \,\, e^{ - D_{\rm a} / D_{\rm char} } \,\,\,{\rm for \, D_{\rm min} \le D_{\rm a} \le D_{\rm max} }
\label{eqn:ExponentialPSD01}
\end{equation}
\noindent where $N_0$ is a normalization constant, $D_{\rm char}$ is a characteristic size, and the PSD is assumed to be truncated below $D_{\rm a} = D_{\rm min}$ and above $D_{\rm a} = D_{\rm max}$. Substituting Eqn.\,\ref{eqn:ExponentialPSD01} into Eqn.\,\ref{eqn:OpticalDepthIntegral01} and using Eqn.\,\ref{eqn:ExtinctionCrossSection01} gives: 
%
\begin{equation}
N_0 ~= ~\frac{ 2 \, \, \tau_{\rm los} }{ \pi \, D_{\rm char} \,
       ( {\rm t_1} - {\rm t_2} ) }
\label{eqn:ExponentialPSD02}
\end{equation}
\noindent where
%
\begin{equation}
{\rm t_1} ~= ~( 2 D_{\rm char}^2 + 2 D_{\rm char} D_{\rm min} + D_{\rm min}^2 )
            \, e^{- D_{\rm min} / D_{\rm char} } 
\label{eqn:ExponentialPSD03}
\end{equation}
\noindent and
%
\begin{equation}
{\rm t_2} ~= ~( 2 D_{\rm char}^2 + 2 D_{\rm char} D_{\rm max} + D_{\rm max}^2 )
            \, e^{- D_{\rm max} / D_{\rm char} } 
\label{eqn:ExponentialPSD04}
\end{equation}
\noindent Analytic forms for $P_{\rm apx}(\theta)$ are even more complicated than those using the power-law form and are not investigated in this work. However, it is useful to note that since the exponential form tends to fall off very rapidly for $D_{\rm a} \gg D_{\rm char}$, frequently little is lost by taking $D_{\rm max} \rightarrow \infty$. As a result, the number of free parameters tends to be 1 fewer than for the power-law PSD. 

\begin{figure}
\noindent\includegraphics[width=20pc]{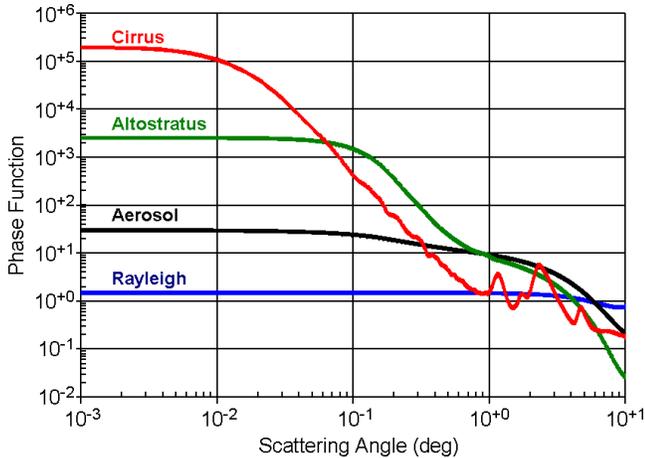}
\caption{The phase functions used to determine the importance of multiple scattering.}
\label{fig:F11_MulitpleScatteringPhaseFunctions}
\end{figure}

\subsubsection{Total Phase Function Approximation}
\label{subsub:TotalPhaseFunctionApproximation}

\par The example phase functions calculated for power-law PSDs presented above [Fig.\,\ref{fig:F10_PowerLawPhaseFunctions} (a)] and the analytic solution using $P_{\rm bp}(\theta,D_{\rm a})$ (Eqns.\,\ref{eqn:Broken04} and \ref{eqn:Broken07}) suggest that an analytic function $P_{\rm apx}(\theta)$ representing $P(\theta)$ in the first two regions, i.e., for scattering angles before the $\theta^{-3}$ behavior sets in, might be approximated by replacing the $\theta^3$ term with $\theta^\nu$ in the denominator of the approximate individual particle phase function $P_{\rm apx}(\theta,D_{\rm a})$:  
%
\begin{equation}
P_{\rm apx}(\theta) ~= ~\frac{ P_0 }{ 1 + ( \theta / \theta_0 )^\nu }
\label{eqn:AnalyticApproximation05}
\end{equation}
\noindent where $P_0$, $\theta_0$, and $0 \lesssim \nu \lesssim 3$ are constants to be determined from fitting either calculations or measurement data. In fact, based on our results for the power-law PSD, we might expect $\nu \simeq 5 - \mu$. Fig.\,\ref{fig:F10_PowerLawPhaseFunctions} (b) compares four of the calculations of $P(\theta)$ with fits, $P_{\rm apx}(\theta)$. The fits were carried out using the Levenberg-Marquardt method (Press et al. 1992) for values of $\theta$ below where the curves appear to start bending over to the $\theta^{-3}$ region. The closeness of the fits confirms the potential utility of this functional form.

\subsection{Radiance Profiles}
\label{sub:RadianceProfiles}

\par The aureole radiance $L_{ss}(\theta)$ resulting from the single scattering of starlight by ice crystals uniformly distributed in a plane-parallel layer depends upon their total phase function $P(\theta)$ as follows (e.g., pg. 302, Liou, 2002): 
%
\begin{equation}
L_{ss}(\theta) ~= ~\frac{ \tau_{\rm los} ~e^{-\tau_{\rm los}} ~S_0(\lambda) }{ 4 \pi } ~P(\theta)
\label{eqn:AureoleRadiance01}
\end{equation}
\noindent where the single scattering albedo (the ratio of the scattering to the extinction cross section) has been taken as 1, $S_0(\lambda)$ is the exo-atmospheric irradiance of the star, and the small-angle approximation has been applied. When Eqn.\,\ref{eqn:AureoleRadiance01} applies, it is a relatively simple matter to solve for $P(\theta)$ given measurements of the aureole radiance and $\tau_{\rm los}$, and using a previously determined value of $S_0(\lambda)$. However, numerous authors (e.g., Dave, 1964; Hovenier, 1971; Korkin et al., 2012) have commented on the limited applicability of the single scattering approximation, e.g., to $\tau_{\rm los} \lesssim 0.05$. These studies, however, typically considered scattering from small particles at all angles. The situation is distinctly different for scattering from large particles at small angles as discussed next. 

\subsubsection{Multiple Scattering}
\label{subsub:MultipleScattering}

\par To investigate the limits of the applicability of the single scattering approximation for use in modeling aureole radiance profiles, we used four phase functions representing the full range of scattering angles. Starting with the least forward scatterering particles, we looked at Rayleigh scattering, which has a simple, analytic phase function. As an example of aerosols we used the phase function retrieved by AERONET (Holben et al., 1998) at NASA Goddard Space Flight Center at 18.636 UT on 8 January 2010. We calculated a phase function for altostratus using the PSD shown by Liou (2002) and the Mie code of Bohren and Huffman (1983). For an example of cirrus we used the phase function calculated by Baum et al. (2005) for a distribution of ice crystal habits and sizes with effective size of $80 \,{\rm \mu m}$. Fig. \ref{fig:F11_MulitpleScatteringPhaseFunctions} plots these phase functions on a log-log scale to emphasize the forward scattering direction. This clearly demonstrates the dominance of forward scattering for large particles. 

\begin{figure}
\noindent\includegraphics[width=20pc]{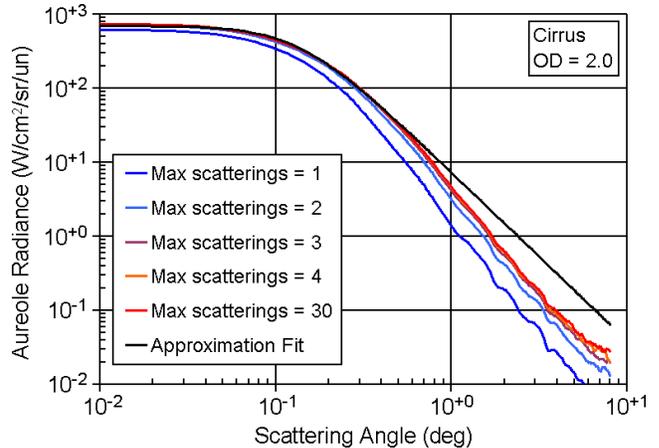}
\caption{Monte Carlo calculations of the aureole radiance profiles for cirrus with $\tau_{\rm los} = 2.0$ (``OD = 2.0'' in the figure legend) including 1, 2, 3, 4, and 30 scatterings and a fit using Eqn.\,\ref{eqn:AnalyticApproximation06}.}
\label{fig:F12_CirrusAureoleRadianceCalculations}
\end{figure}

\begin{figure*}
\begin{center}
\includegraphics[width=39pc]{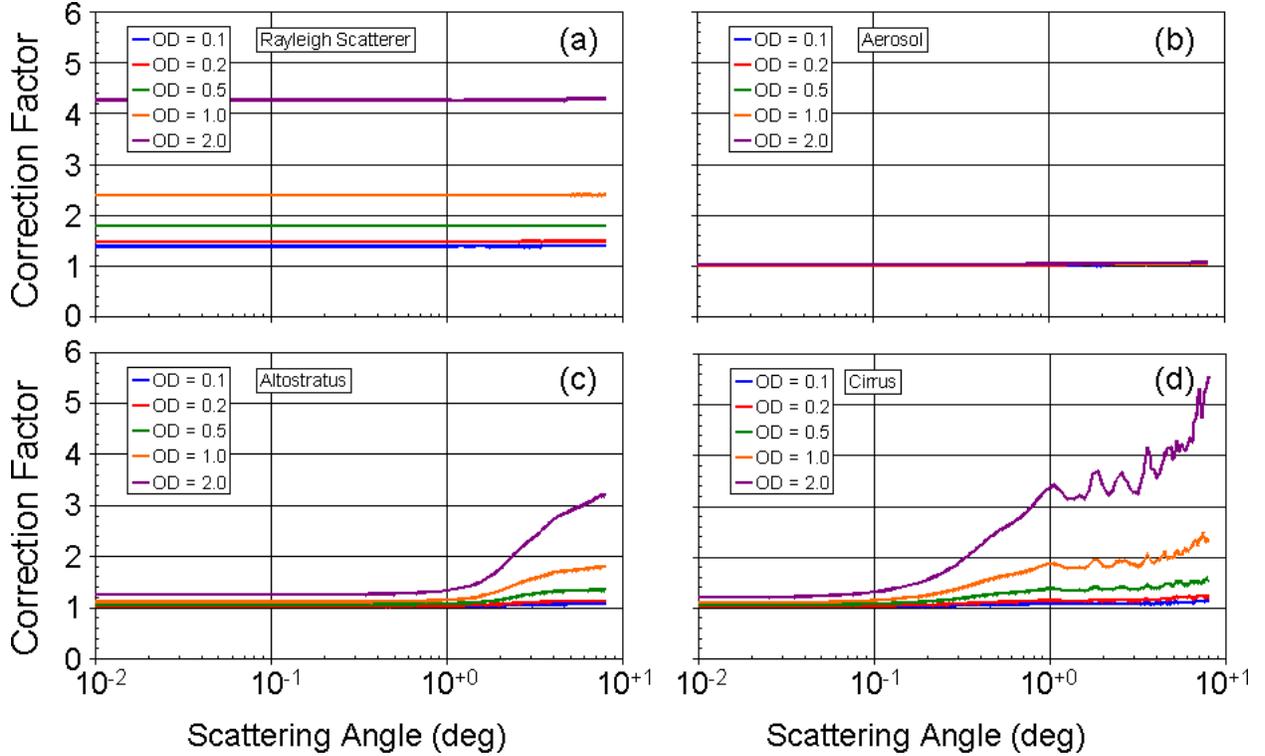}
\caption{Multiple scattering scattering correction factor for (a) Rayleigh scatterers, (b) aerosols, (c) altostratus, and (d) cirrus for 5 values of $\tau_{\rm los}$ (``OD'' in the figure legend) from $0.1$ to $2.0$.}
\label{fig:F13_MultipleScatteringCorrectionFactors}
\end{center}
\end{figure*}

\par We digress briefly to describe the radiative transfer (RT) method we used to calculate the scattering of starlight through a uniform, plane-parallel particulate layer. The strongly forward peaked phase functions characteristic of ice crystals in cirrus (Fig.\,\ref{fig:F11_MulitpleScatteringPhaseFunctions}) tend to cause problems for many RT algorithms. Methods relying on truncating the peak of the phase function at small angles (for example, Joseph et al., 1976; Wiscombe, 1977) work well for hemispheric flux calculations, but are not concerned with details of the angular distribution of radiation and therefore are not appropriate for calculating the profiles of stellar aureole radiance scattered at small angles. We used the successive orders of scattering method (Evans and Marshak, 2005), where the integrals were calculated using a Monte Carlo method (Sobol, 1994). In order to make the calculations more efficient we solved the adjoint problem, tracing photons from the sensor back to the source. Fig.\,\ref{fig:F12_CirrusAureoleRadianceCalculations} shows an example of calculations for cirrus with $\tau_{\rm los} = 2.0$. The exo-atmospheric stellar irradiance was taken as that of the Sun ($S_0 = 0.1475 \,{\rm W \, cm^{-2} sr^{-1} \mu m^{-1}}$) for these calculations. The colored lines labelled ``Max Scatterings ='' show calculations where the maximum number of scatterings was limited to 1, 2, 3, 4, and 30. (The black curve labelled ``Approximation Fit'' is discussed later.) The dark blue curve shows the single scattering solution. The fine structure for $\theta \gtrsim 1^\circ$ is numerical noise from the Monte Carlo calculations. This noise decreases somewhat as the maximum number of scatterings increases. The radiance in this region falls off as $\approx \theta^{-3}$. As expected, the aureole broadens somewhat as the maximum number of scatterings increases. For the range of angles shown, the solution appears to have converged using a maximum of three or four scatterings.


\par We define a multiple scattering correction factor as the ratio of the aureole radiance calculated using a maximum of 30 scatterings to that calculated for single scattering. Fig.\,\ref{fig:F13_MultipleScatteringCorrectionFactors} shows calculations of the correction factors using the phase functions shown in Fig.\,\ref{fig:F11_MulitpleScatteringPhaseFunctions} for $\tau_{\rm los}$ = 0.1, 0.2, 0.5, 1.0, and 2.0. The Rayleigh scattering case (panel a) shows a significant increase in radiance as the maximum number of scatterings increases, and has not converged by four scatterings (not shown). The correction factor is nearly constant with $\theta$, reflecting the shape of the phase function. The aerosol case (panel b) is interesting in that although the phase function is much flatter than that of either altostratus or cirrus, the single scattering solution appears adequate. Unlike the other three cases, which represent conservative scattering, this aerosol case is highly absorptive. The single scattering albedo is only 0.12, which inhibits multiple scattering significantly. The correction factors for the altostatus (panel c) and cirrus (panel d) are close to 1 or nearly constant for $\theta$ less than $\sim$$1^\circ$ and $\sim$$0.1^\circ$, respectively, meaning that the shapes of the aureoles differ only slightly from those of the underlying phase functions.

\par Notwithstanding the fact that there is typically relatively little difference between single and multiple scattering in the stellar aureoles from cirrus clouds, it is worthwhile to explore correction algorithms, especially for the higher optical depth cases. And it appears appropriate to try to correct for multiple scattering in these cases, especially at the larger angles. The appendix presents a technique for deconvolving the effects of multiple scattering. 

\begin{figure}
\noindent\includegraphics[width=20pc]{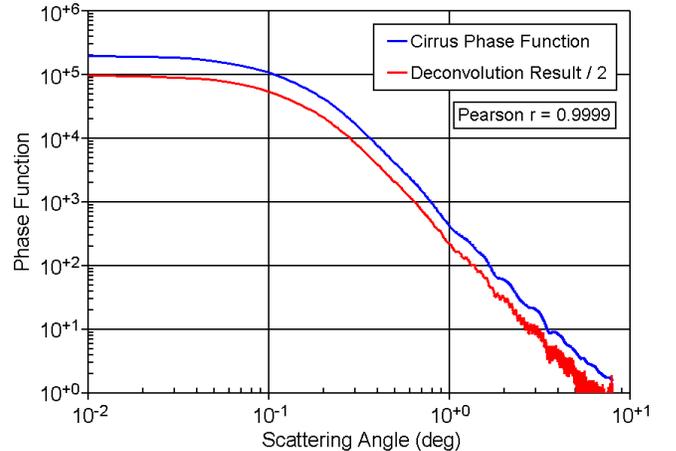}
\caption{Comparison of the phase function (Fig.\,\ref{fig:F11_MulitpleScatteringPhaseFunctions}) used to calculate the aureole radiance for cirrus with $\tau_{\rm los} = 2.0$ (Fig.\,\ref{fig:F12_CirrusAureoleRadianceCalculations}) with the phase function retrieved from the diffraction radiance profile using Eqn.\,\ref{eqn:RetrievedPhaseFunction01} and divided by 2 for clarity. }
\label{fig:F14_DeconvolutionExample}
\end{figure}

\begin{figure*}
\begin{center}
\includegraphics[width=39pc]{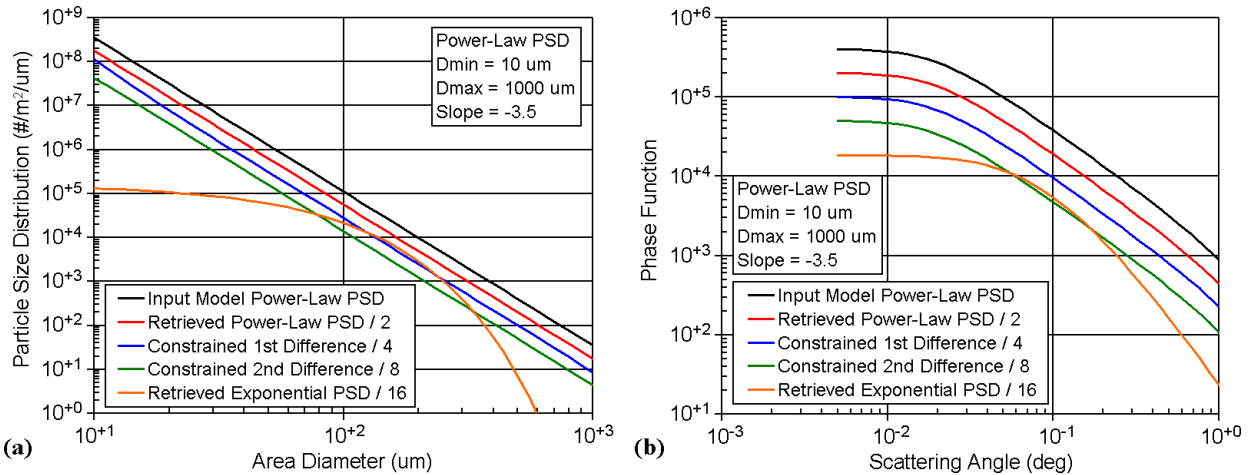}
\caption{ (a) Comparison of PSDs retrieved by fitting power-law and exponential PSDs and by application of numerical inversion using first and second difference constraints to the phase function generated from an input model power-law PSD. (b) Comparison of phase functions calculated numerically from the retrieved PSDs with the input model phase function. To separate the curves four of them have been reduced by a factor as indicated in the figure legends. }
\label{fig:F15_RetrievalExample}
\end{center}
\end{figure*}


\subsubsection{Aureole Approximation}
\label{subsub:AureoleApproximation}

\par To the extent that the shapes of aureoles differ only slightly from those of the underlying phase functions, it is reasonable to assume that an analytic approximation similar to the one found for phase functions could be useful in the near forward direction. Consider the aureole profile approximation $L_{\rm apx}(\theta)$ based on Eqn.\,\ref{eqn:AnalyticApproximation05}:  
%
\begin{equation}
L_{\rm apx}(\theta) ~= ~\frac{ L_0 }{ 1 + ( \theta / \theta_0 )^\nu }
\label{eqn:AnalyticApproximation06}
\end{equation}
\noindent where $L_0$, $\theta_0$, and $\nu$ are constants to be determined from fitting either model calculations or measurement data. 

\par As an example, the black line in  Fig.\,\ref{fig:F12_CirrusAureoleRadianceCalculations} shows an illustrative  fit to the red curve representing the multiple scattering aureole radiance profile for a cirrus cloud with $\tau_{\rm los} = 2.0$. The fit for $\theta \lesssim 0.5^\circ$ is obviously very good.  

\subsubsection{Phase Function Retrieval}
\label{subsub:PhaseFunctionRetrieval}

\par The phase function $P(\theta)$ can be retrieved or deconvolved from a measured diffraction radiance profile $L_{\rm dif}(\theta)$ numerically using the Hankel transform solution described in \S \ref{sub:SeriesSolution}, specifically Eqns.\,\ref{eqn:SeriesSolution06} and \,\ref{eqn:TotalScatteringAureole01}:
%
\begin{equation} 
P(\theta) ~= ~\frac{ 4 \pi }{ \tau_{\rm los} } ~\mathcal{H} \left \{ \ln \left ( 1 + e^{\tau_{\rm los}} \mathcal{H} \left \{ \frac{ L_{\rm dif}(\theta) }{ S_0 } \right \} \right ) \right \}
\label{eqn:RetrievedPhaseFunction01}
\end{equation}
\noindent where $S_0$ is the exo-atmospheric stellar irradiance and $\mathcal{H} \{ \}$ is the Hankel transform (Eqn.\,\ref{eqn:Hankel01}). 

\par Fig.\,\ref{fig:F14_DeconvolutionExample} compares the phase function (the blue curve) used to calculate the aureole through a cirrus cloud with $\tau_{\rm los} = 2.0$ (shown in Fig.\,\ref{fig:F12_CirrusAureoleRadianceCalculations}) with a retrieval (the red curve) using Eqn.\,\ref{eqn:RetrievedPhaseFunction01}. The two curves overlap so well that we have divided the latter curve by a factor of 2 so that it can be distinguished in the plot. The deconvolution result exhibits numerical noise for $\theta \gtrsim 1^\circ$, which some simple smoothing could mitigate. To the best of our knowledge this deconvolution technique is new, and the effects of measurement errors and noise have yet to be worked out. The result shown in Fig.\,\ref{fig:F14_DeconvolutionExample} lacks such noise and serves simply to confirm that this technique can work under ideal conditions. The reasonableness of the results presented later in the paper for actual stellar aureole data serve to add confidence in the technique under more realistic conditions. Although RT calculations could be used to check whether a retrieved phase function reproduces the measured aureole profiles, we prefer to perform such checks starting with the PSDs derived from the phase functions and seeing how well the phase functions are reproduced.

\subsection{PSD Retrieval}
\label{sub:PsdRetrieval}

\par Given the phase function $P(\theta)$ and the optical depth $\tau_{\rm los}$ there are several ways of attempting to retrieve the PSD. We consider two complementary techniques.

\subsubsection{Fitting an Analytic Form}
\label{subsub:FittingAnAnalyticForm}

\par In \S \ref{subsub:PowerLawDistribution} and \S  \ref{subsub:ExponentialDistribution} two parametric approximations were shown for representing PSDs. Given $\tau_{\rm los}$, the power-law PSD has three free parameters ($\mu$, $D_{\rm min}$, and $D_{\rm max}$) and the exponential PSD two ($D_{\rm char}$ and $D_{\rm min}$, with $D_{\rm max}$ set to $1000 \,{\rm \mu m}$). A $\chi^2$ error metric can be defined as:
%
\begin{equation} 
\chi^2 ({\bf p}) ~= ~\sum_{i=1}^n ~w_i^2 ~[ P_{\rm data}(\theta_i) - P_{\rm model}(\theta_i,{\bf p}) ]^2
\label{eqn:ErrorMetric01}
\end{equation}
\noindent where ${\bf p}$ is a vector of the free parameters, $P_{\rm data}(\theta_i)$ are the values of the phase function derived (deconvolved) from the aureole measurements, $w_i$ are suitable weights, e.g., setting $w_i = 100/P_{\rm data}^2(\theta_i)$ is equivalent to assuming $10\%$ errors, and $P_{\rm model}(\theta_i,{\bf p})$ are the values of the phase function calculated using the model PSD with parameters ${\bf p}$ and Eqns.\,\ref{eqn:AnalyticApproximation04} to \ref{eqn:OpticalDepthIntegral01}. The Levenberg-Marquardt method (e.g., Press et al. 1992) can be applied to minimize $\chi^2 ({\bf p})$ with the required derivatives calculated numerically. 


\par As an example fit consider again a model power-law PSD extending from $D_{\rm min} = 10 \,{\rm \mu m}$ to $D_{\rm max} = 1000 \,{\rm \mu m}$ with $\mu = 3.5$. The plot in Fig.\,\ref{fig:F15_RetrievalExample} (a) compares the input model PSD (the black curve) used to generate the synthetic aureole with retrievals using analytic models involving the power-law and exponential forms. (The constrained retrievals are discussed below.) Not surprisingly the retrieval using the power-law form (the red curve) does quite well, while the retrieval using the exponential form (the orange curve) is only close over part of the size range. As with the multiple scattering deconvolution (\S \ref{subsub:AureoleApproximation}) these calculations lack noise and serve simply to confirm that the technique can work under ideal conditions and assuming that the correct underlying form of the PSD is used. Fig.\,\ref{fig:F15_RetrievalExample} (b) compares the phase functions calculated numerically using the retrieved PSDs shown in panel (a). The fact that the phase function calculated from the retrieved PSD when an exponential form is used does not match the input PSD in this case reassures us that any arbitrary form cannot be assumed. In principle then a variety of PSD models could be used and the one producing the best fit to the input phase function selected. However, the next section presents an alternative approach that has worked well for a variety of other physical retrieval problems.

\subsubsection{Constrained Numerical Inversion}
\label{subsub:ConstrainedNumericalInversion}

\par To avoid assuming a specific functional form for the PSD as in the previous section we consider the more direct numerical solution of Eqn.\,\ref{eqn:AveragePhaseFunction01}. The numerical solution of this Fredholm integral equation for $N(D_{\rm a})$ as a function of $P(\theta)$  can be made easier (King et al., 1978) by replacing the independent variable $N(D_{\rm a})$ with the product of a slowly varying function, $f(D_{\rm a})$, and a more rapidly varying one, e.g., $D_{\rm a}^{-\beta}$: 
%
\begin{equation}
N(D_{\rm a}) ~\equiv ~f(D_{\rm a}) ~D_{\rm a}^{-\beta}
\label{eqn:ModifiedVariable01}
\end{equation}
\noindent where $\beta = 4$ was suggested by the large-particle power-law slopes found by Heymsfield and Platt (1984) and seems to work reasonably well for the current problem. Substituting Eqn.\,\ref{eqn:ModifiedVariable01} into Eqn.\,\ref{eqn:AveragePhaseFunction01} gives:
%
\begin{equation}
P(\theta)~= ~ 
\frac{ \int_0^\infty \sigma_{\rm ext}(D_{\rm a}) ~P_{\rm apx}(\theta,D_{\rm a}) ~D_{\rm a}^{-\beta}~ f(D_{\rm a}) ~d D_{\rm a} }{ \tau_{\rm los} }
\label{eqn:AveragePhaseFunction04}
\end{equation}
\par Following the usual inversion approach (for example, Twomey, 1977), we discretize this equation as follows:
%
\label{eqn:Discretization01}
\begin{equation}
\begin{split}
P(\theta_i) & ~= ~ \label{eqn:Discretization01} \\
\sum_{j=1}^m &~\int_{D_{j-\frac{1}{2}}}^{D_{j+\frac{1}{2}}} ~\frac{ \sigma_{\rm ext}(D') ~P_{\rm apx}(\theta_i,D') ~D'^{-\beta} }{ \tau_{\rm los} } ~d D' ~f(D_j) 
\end{split}
\end{equation}
\noindent where the $D_j$ form a discrete sequence of $m$ values spanning the range of area diameters from $D_{\rm min}$ to $D_{\rm max}$,  $D_{j\pm\frac{1}{2}}$ is suitably defined, e.g., $D_{j \pm  \frac{1}{2}} = ( D_j + D_{j\pm 1} ) / 2$ if the values are linearly spaced, and $f(D_j)D_j^{-\beta}$ is the average particle density in the size interval about $D_j$. It is conventional to express Eqn. \ref{eqn:Discretization01} using matix notation:
%
\begin{equation}
\bf{g} ~= ~\bf{A} ~\bf{f}
\label{eqn:Discretization02}
\end{equation}
\noindent where 
%
\begin{equation}
g_i ~= ~P(\theta_i)
\label{eqn:Discretization03}
\end{equation}
%
\begin{equation}
f_j ~= ~f(D_j)
\label{eqn:Discretization04}
\end{equation}
\noindent and
%
\begin{equation}
A_{i,j} ~= ~\int_{D_{j-\frac{1}{2}}}^{D_{j+\frac{1}{2}}} ~\frac{ \sigma_{\rm ext}(D') ~P_{\rm apx}(\theta_i,D') ~D'^{-\beta} }{ \tau_{\rm los} } ~d D'
\label{eqn:Discretization05}
\end{equation}
\noindent Substituting from Eqns. \ref{eqn:AnalyticApproximation04} and \ref{eqn:ExtinctionCrossSection01} we find:
%
\begin{equation}
A_{i,j} ~= ~\frac{ \pi^3 }{ 4 ~\lambda^2 ~\tau_{\rm los} } ~\int_{D_{j-\frac{1}{2}}}^{D_{j+\frac{1}{2}}} ~\frac{ D'^{4-\beta} }{ 1 + ( \pi \xi \theta D' / \lambda )^3 } ~d D'
\label{eqn:Discretization06}
\end{equation}
\noindent where $A_{i,j}$ is calculated approximately and analytically by assuming a power-law dependence of the integrand between the end points of each particle size interval.

\begin{figure*}
\begin{center}
\includegraphics[width=39pc]{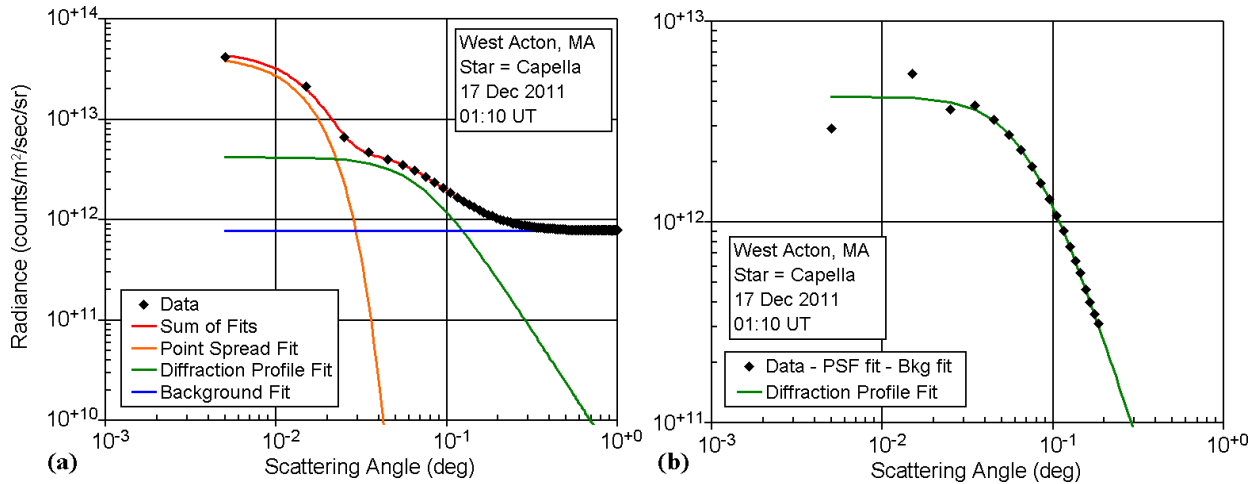}
\caption{(a) Analysis of the components comprising the aureole profile around Capella at 01:10 UT on 17 Dec 2011 and (b) comparison of the diffraction profile fit with the data minus the PSF and background fits.}
\label{fig:F16_AureoleComponents}
\end{center}
\end{figure*}

\par In general, $\bf{A}$ is not square and hence Eqn. \ref{eqn:Discretization02} cannot be inverted directly. The least squares solution to this equation, $\bf{f} = ( \bf{A}^T \bf{A} )^{-1} ~\bf{A}^T ~\bf{g}$, is problematic because it is under constrained (Twomey, 1977). Frequently, a practical solution makes use of the addition of an appropriate constraint, e.g., that the solution be smooth in some sense. The constraint takes the form of a second equation that is fit simultaneously. A Lagrange multiplier $\Lambda$ is selected to govern the balance between the least squares errors of the two equations:  
%
\begin{equation}
\bf{f} ~= ~( \bf{A}^T \bf{A} + \Lambda \bf{H} )^{-1} \bf{A}^T \bf{g}
\label{eqn:ConstrainedEquation01}
\end{equation}
\noindent To look for a smooth solution, one can constrain (minimize) the first differences by taking $\bf{H}$ as:
%
\begin{equation}
\left[ \begin{array}{rrrrrrrrr} 
  1 & -1 &  0 &  0 &  0 &  0 &... \\ 
 -1 &  2 & -1 &  0 &  0 &  0 &... \\ 
  0 & -1 &  2 & -1 &  0 &  0 &... \\ 
~~. &~~. &~~. &~~. &~~. &~~. &... \\
 ...&  0 &  0 & -1 &  2 & -1 &  0 \\
 ...&  0 &  0 &  0 & -1 &  2 & -1 \\
 ...&  0 &  0 &  0 &  0 & -1 &  1 
\end{array} \right]
\label{eqn:ConstrainedEquation02}
\end{equation}
\noindent or the second differences by taking $\bf{H}$ as:
%
\begin{equation}
\left[ \begin{array}{rrrrrrrrr} 
  1 & -2 &  1 &  0 &  0 &  0 &  0 &... \\ 
 -2 &  5 & -4 &  1 &  0 &  0 &  0 &... \\ 
  1 & -4 &  6 & -4 &  1 &  0 &  0 &... \\ 
  0 &  1 & -4 &  6 & -4 &  1 &  0 &... \\ 
~~. &~~. &~~. &~~. &~~. &~~. &~~. &... \\
 ...&  0 &  1 & -4 &  6 & -4 &  1 &  0 \\
 ...&  0 &  0 &  1 & -4 &  6 & -4 &  1 \\
 ...&  0 &  0 &  0 &  1 & -4 &  5 & -2 \\
 ...&  0 &  0 &  0 &  0 &  1 & -2 &  1 
\end{array} \right]
\label{eqn:ConstrainedEquation03}
\end{equation}
\noindent Press et al. (1992) recommend the following choice for $\Lambda$, which  balances the two components of the minimization:
%
\begin{equation}
\Lambda ~= ~\frac{ Tr( \bf{A}^T \bf{A} ) }{ Tr( \bf{H} ) }
\label{eqn:ConstrainedEquation04}
\end{equation}
\noindent where $Tr()$ is the trace of the matrix. 

\par Fig. \ref{fig:F15_RetrievalExample} also shows the results of constrained inversions. Both constraints work reasonably well for the power-law model case, although the second difference did marginally better than the first difference constraint. 

\subsection{Point Spread Function}
\label{sub:PointSpreadFunction}

\par We model stellar aureoles as the sum of the instrumental/atmospheric point spread function (PSF), starlight diffraction, and background, e.g., diffusely illuminated sky. The PSF describes the spreading of the radiation from a point source, such as a star, and results from a combination of physical processes, e.g., diffraction by optical elements, saturation of detectors, bleeding of charge on the focal plane, and spreading by atmospheric turbulence. Since the latter process tends to be variable with time and space, we approach modeling the PSF using an empirical parameterization. The Gaussian function $G(\theta)$ is frequently used to model PSFs (Racine, 1996):
%
\begin{equation}
G(\theta) ~= ~g_0 \, e^{ - \frac{ \theta^2 }{ 2 \theta_g^2 } }
\label{eqn:Gaussian01}
\end{equation}
\noindent where $g_0$ and $\theta_g$ are fitting parameters. The effects of atmospheric turbulence appear to be modeled somewhat better using the Moffat function $M(\theta)$ (e.g., Moffat, 1969; Trujillo et al. 2001):
%
\begin{equation}
M(\theta) ~= ~\frac{ m_0 }{ ( \, 1 + ( \theta / \theta_m )^2 \, )^{m_1} }
\label{eqn:Moffat01}
\end{equation}
\noindent where $m_0$, $\theta_m$, and $m_1$ are fitting parameters. However, in this work we have found the Gaussian function preferable because its functional form is more distinct from that of the aureole approximation (Eqn.\,\ref{eqn:AnalyticApproximation06}). This difference allows the two functional forms to be fit simultaneously without their roles being confused in the process. Therefore we have adopted the Gaussian functional form for modeling PSFs. 

\begin{figure}
\noindent\includegraphics[width=20pc]{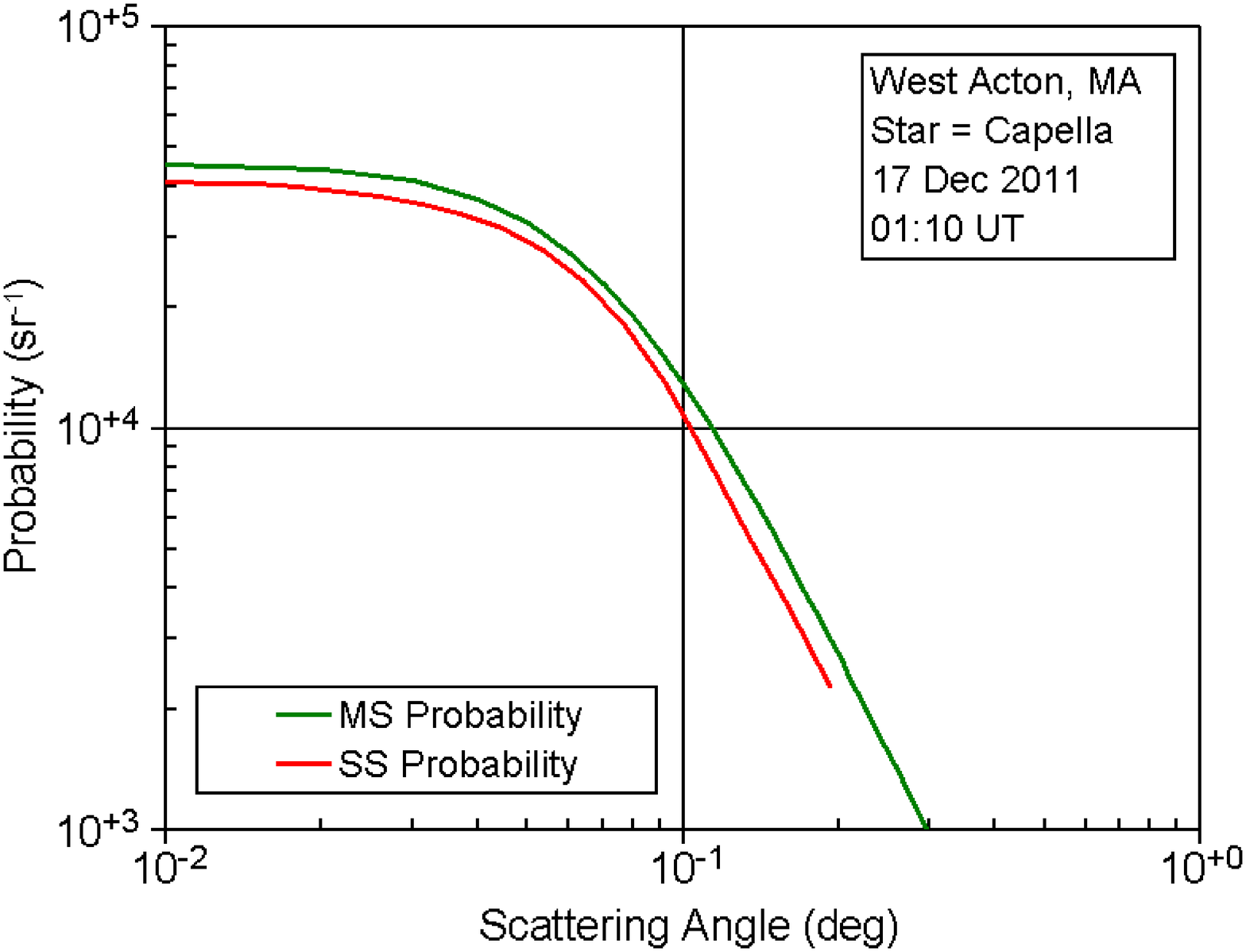}
\caption{The green curve shows the multiple scattering probability per unit solid angle, $Q_{\rm ms}(\theta) = L_{\rm apx}(\theta) / S_0$, calculated from the diffraction profile fit shown in Fig.\,\ref{fig:F16_AureoleComponents} (b). The red curve shows the single scattering probability per steradian, $Q(\theta)$, deconvolved from $Q_{\rm ms}(\theta)$ using Eqn.\,\ref{eqn:SeriesSolution05}.}
\label{fig:F17_AureoleDeconvolution}
\end{figure}

\section{Aureole Profile Interpretation}
\label{sec:AureoleProfileInterpretation}

\par The process of interpreting aureole profiles begins with separating the three major physical components, the PSF, the stellar diffraction profile, and the background. Next the stellar diffraction profile is deconvolved (for the effects of multiple scattering) to give the phase function, from which the PSD is then derived. The process is illustrated in this section using aureole data for Capella from about 01:10 UT on 17 Dec 2011. 

\subsection{Diffraction Profile Extraction}
\label{sub:DiffractionProfileExtraction}

\par The center of Capella on the image was identified, first by locating the brightest pixel and then refining the location using the maximum of a quadratic surface fit to the pixels within $0.1^\circ$ of the brightest point. A set of concentric annuli was defined, and the pixel values within each annulus were averaged to produce the aureole radiance values shown as the black diamonds in Fig.\,\ref{fig:F16_AureoleComponents} (a).

\par Using Eqn.\,\ref{eqn:Gaussian01} for the PSF, Eqn.\,\ref{eqn:AnalyticApproximation06} for the stellar aureole, and a constant, $L_{\rm bkg}$, for the sky background, the sum of these components, $L_{\rm aur}(\theta)$, is modeled as:
%
\begin{equation}
L_{\rm aur}(\theta) ~= ~g_0 ~e^\frac{ - \theta^2 }{ 2 \theta_g^2 } + ~\frac{ L_0 }{ 1 + ( \theta / \theta_0 )^\nu } + L_{\rm bkg}
\label{eqn:AureoleProfile01}
\end{equation}

\noindent We find the 6 parameters, $g_0$, $\theta_g$, $L_0$, $\theta_0$, $\nu$, and $L_{\rm bkg}$ using the Levenberg-Marquardt method to fit the data points. We tried using the formal RMS flucutations of the pixel values in each annulus to specify the weights used in the fitting process, but this produced error bars that were manifestly smaller than the actual error bars. Consequently, we somewhat arbitrarily adopted $10\%$ error bars for all of the data points to take into account other, systematic errors. We found empirically that the specific choice of weighting functions had relatively little effect on the retrieved diffraction pattern of the aureole profile.  Fig.\,\ref{fig:F16_AureoleComponents} (b) compares the fit for the stellar aureole with data points after subtracting off the fitted PSF and background values. The plot suggests that the aureole approximation $L_{\rm apx}(\theta)$ (Eqn.\,\ref{eqn:AnalyticApproximation06}) has done a reasonable job of representing the diffraction of starlight by the cirrus cloud particles.  

\begin{figure*}
\begin{center}
\includegraphics[width=39pc]{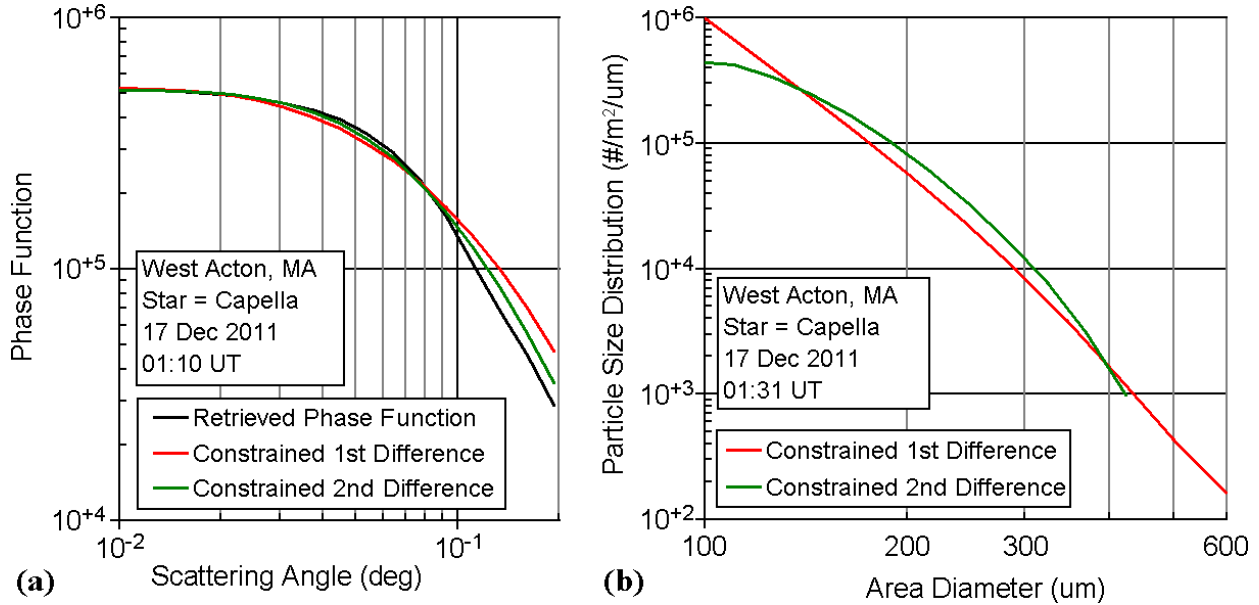}
\caption{ The black curve in panel (a) shows the retrieved phase function calculated by multiplying the single scattering probability shown in Fig.\,\ref{fig:F17_AureoleDeconvolution} (the green curve) by $4 \pi$. The red and green curves in panel (b) show the PSDs calculated from this phase function using numerical inversion with first and second difference constraints, respectively. To check the retrieved PSDs, the red and green curves in panel (a) are the phase functions calculated numerically from the corresponding PSDs in panel (b). }
\label{fig:F18_AureoleRetrieval}
\end{center}
\end{figure*}

\subsection{Phase Function Deconvolution}
\label{sub:PhaseFunctionDeconvolution}

\par Next the phase function of the scatterers is derived from the diffraction radiance profile {\it fit}, $L_{\rm apx}(\theta)$, and the exo-atmospheric stellar irradiance, $S_0$, using the deconvolution expression of Eqn.\,\ref{eqn:RetrievedPhaseFunction01}. Although they do not need to be calculated explicitly, it may be helpful to show two of the intermediate products in this process. The green curve in Fig.\,\ref{fig:F17_AureoleDeconvolution} shows the multiple scattering probability per unit solid angle, $Q_{ms}(\theta) \equiv L_{\rm apx}(\theta) / S_0$ (Eqn.\,\ref{eqn:TotalScatteringAureole01}). The red curve shows the single scattering probability per steradian, $Q(\theta)$, deconvolved from $Q_{ms}(\theta)$ using Eqn.\,\ref{eqn:SeriesSolution05}. Note that $Q(\theta)$ is both slightly lower and narrower than $Q_{\rm ms}(\theta)$. This is expected since the effect of multiple scattering is both to increase the total number of photons scattered and to widen their angular distribution. The phase function, the end product of the process, is simply $4 \pi$ times $Q(\theta)$ (Eqn.\,\ref{eqn:SingleScatteringAureole03}).

\subsection{PSD Solution}
\label{sub:PsdSolution}

\par The last step is to retrieve the PSD from the phase function. We use the numerical inversion technique  (\S\,\ref{subsub:ConstrainedNumericalInversion}), trying both first and second difference constraints, and selecting the one that fits $P(\theta)$ better. Fig.\,\ref{fig:F18_AureoleRetrieval} (a) compares the phase functions derived from the fits (the colored lines) with the phase function (the black curve) scaled from $Q(\theta)$ (the red curve in Fig.\,\ref{fig:F17_AureoleDeconvolution}). The numerical inversion using the second difference constraint (the green curve) produces a slightly better fit than does the one using the first difference constraint (the red curve).

\par Fig.\,\ref{fig:F18_AureoleRetrieval} (b) compares the PSDs corresponding to the fits in Fig.\,\ref{fig:F18_AureoleRetrieval} (a). The matrix used in the inversion solution has dimensions $12 \times 12$. Examination of the 12 eigenvalues for the unconstrained problem shows magnitudes ranging from a maximum of $4 \times 10^7$ to a minimum of $5 \times 10^{-9}$, indicative of ill-conditioning. By contrast, when the second difference constraint is added in using the Lagrange multiplier as indicated in Eqn.\,\ref{eqn:ConstrainedEquation04} the eigenvalues range from $4 \times 10^7$ to $6 \times 10^4$, a considerable improvement that enables a solution to be found. Consideration of the first few eigenvalues of the unconstrained problem, $4 \times 10^7$, $7 \times 10^5$, $3 \times 10^4$, $2 \times 10^3$, and $9 \times 10^0$, would suggest that, practically speaking, there are perhaps 2 or maybe 3 free parameters. Note that although these PSDs correspond well to the phase function deconvolved from the aureole radiance measurement we cannot say that they are unique, merely reasonable.  

\begin{figure}
\noindent\includegraphics[width=20pc]{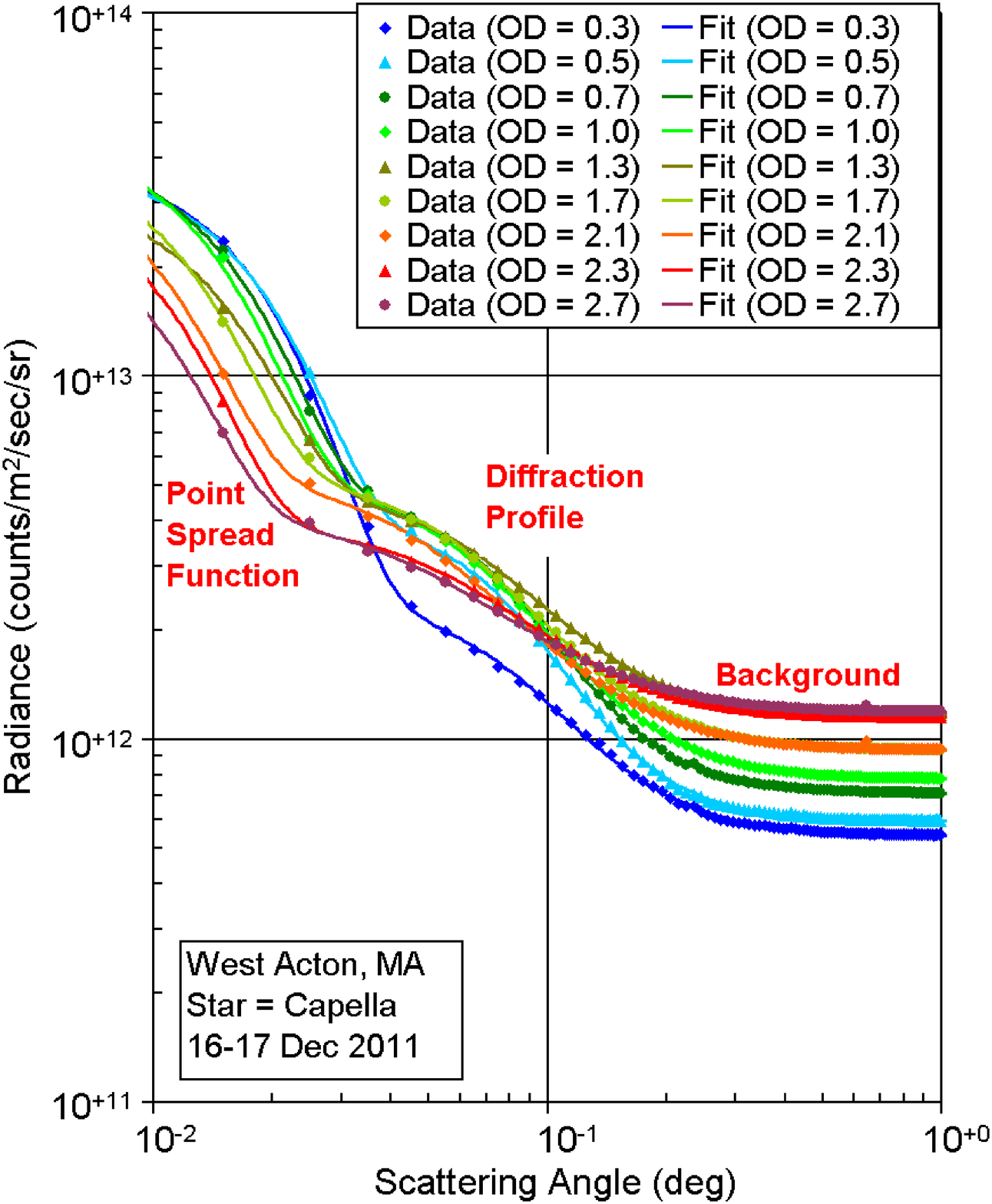}
\caption{Additional examples of aureole profiles and least-squares fits for the star Capella from the night of 16-17 Dec 2011, annotated by $\tau_{\rm los}$ (``OD'' in the figure legend).}
\label{fig:F19_ComponentExamples}
\end{figure}

\begin{figure*}
\begin{center}
\includegraphics[width=39pc]{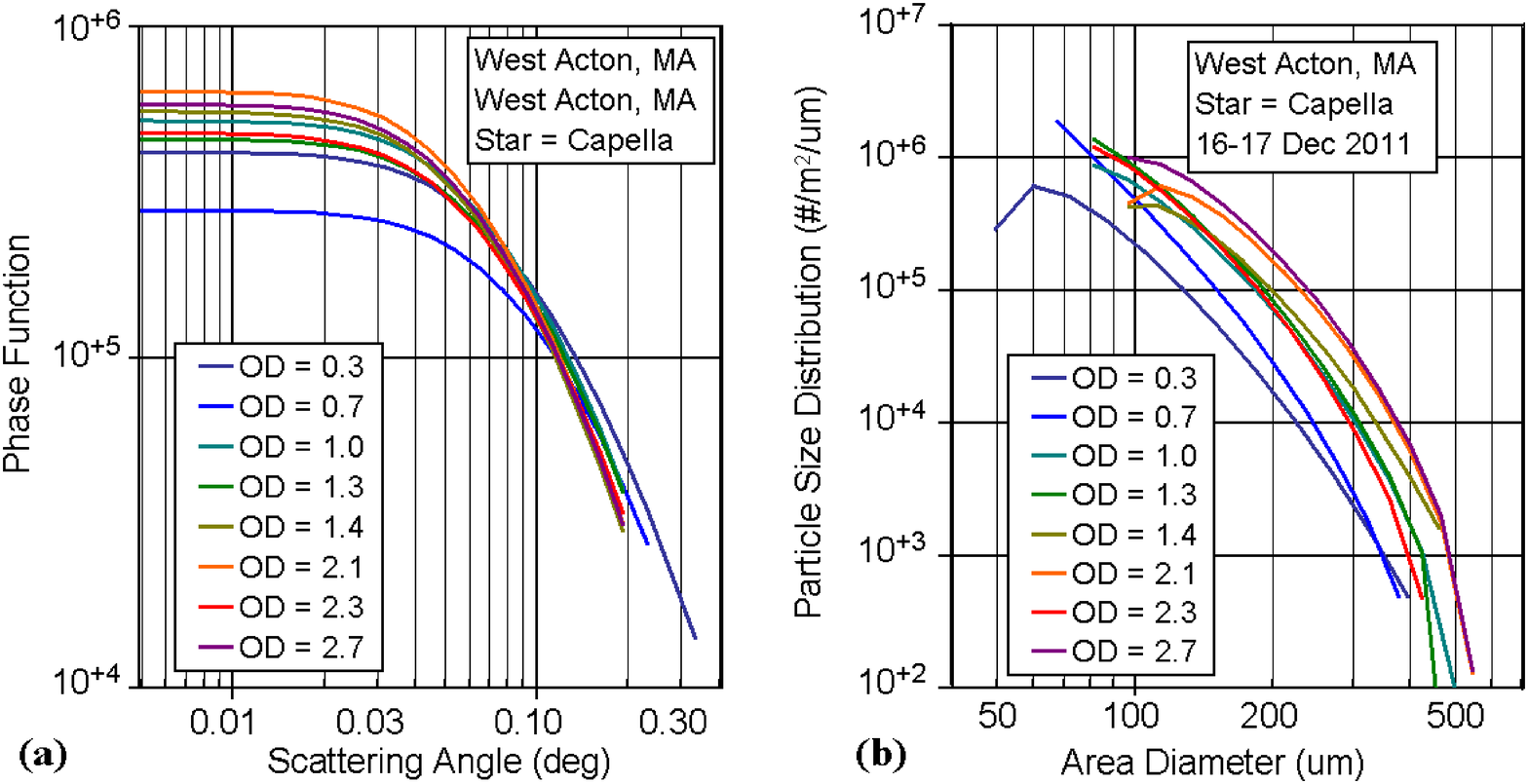}
\caption{(a) Phase functions deconvolved from the diffraction profile fits in Fig.\,\ref{fig:F19_ComponentExamples} and (b) PSD solutions derived from the phase functions using constrained numerical inversion. }
\label{fig:F20_RetrievalExamples}
\end{center}
\end{figure*}

\begin{figure*}
\begin{center}
\includegraphics[width=39pc]{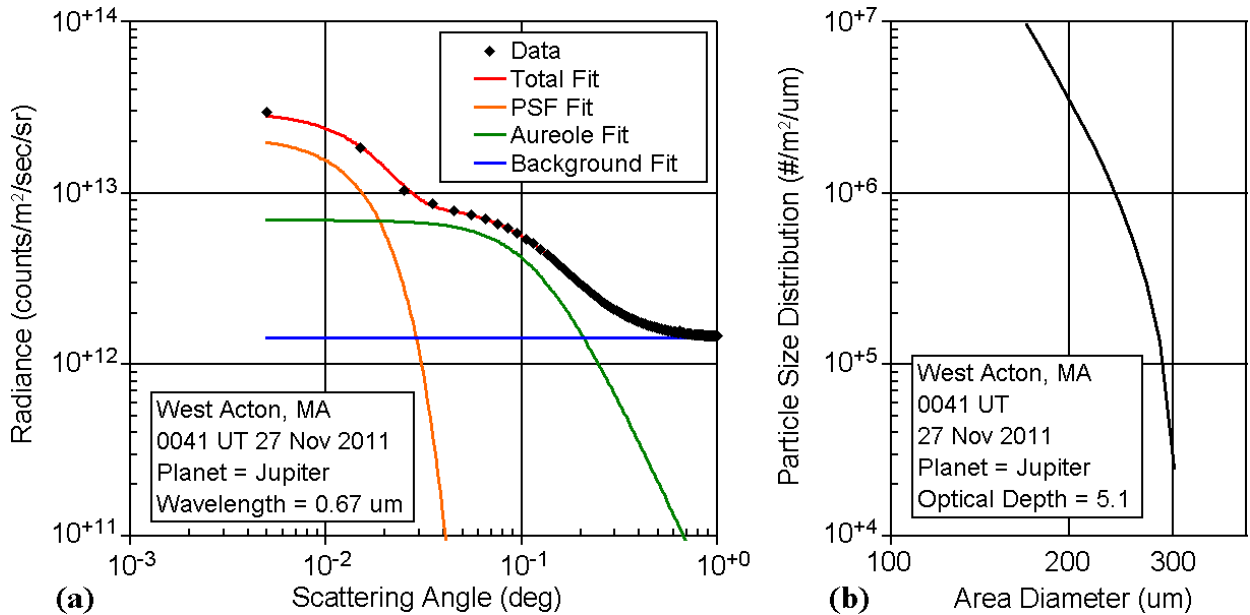}
\caption{(a) Example aureole profile data and least-squares fits for the planet Jupiter at 0041 UT on 27 Nov 2011 when the line-of-sight optical depth was 5.1 and (b) PSD calculated using the second difference constraint.}
\label{fig:F21_JupiterExample}
\end{center}
\end{figure*}

\subsection{More Examples}
\label{sub:MoreExamples}

\par From the datasets we have collected, some 30\% of which have been analyzed in detail, we show a few illustrative examples in this subsection. Fig.\,\ref{fig:F19_ComponentExamples} shows more measurement data from the night of 16-17 Dec 2011. The PSF and background magnitudes are anti-correlated as one might expect, the PSF being brightest when $\tau_{\rm los}$ is least, while the background (resulting largely from backscattered ``cityshine'') is brightest when $\tau_{\rm los}$ is greatest. The diffraction profile radiance is maximized for $\tau_{\rm los} \sim 1$ or so, as is consistent from consideration of the single scattering equation (Eqn.\,\ref{eqn:AureoleRadiance01}).

\par Fig.\,\ref{fig:F20_RetrievalExamples} (a) shows the phase functions deconvolved from the diffraction profile fits in Fig.\,\ref{fig:F19_ComponentExamples}. Interestingly, the magnitude of the phase function forward scattering peak seems to be loosely correlated with $\tau_{\rm los}$. Fig.\,\ref{fig:F20_RetrievalExamples} (b) shows the PSD solutions derived from the phase functions using constrained numerical inversion. The correlation noted above carries through to the PSDs. We are encouraged by the consistency of the retrievals.

\subsubsection{Planetary Aureole Example}
\label{subsub:PlanetaryAureoleExample}

\par Fig.\,\ref{fig:F21_JupiterExample} (a) shows an example of aureole measurement data for the planet Jupiter from the early morning of 27 Nov 2011. Note that the aureole profile does not drop down to the background level until about $0.2^\circ$ in contrast with the case for the star Capella in Fig.\,\ref{fig:F16_AureoleComponents} (a), offering a potentially useful increase in particle measurement range. Nevertheless the PSD retrieved using the second difference constraint [Fig.\,\ref{fig:F21_JupiterExample} (b)] covers roughly the same size range as the retrievals for the star Capella.

\par Examination of the first few eigenvalues of the unconstrainted problem, $2 \times 10^5$, $7 \times 10^2$, and $5 \times 10^0$, suggests that there are perhaps 2 free parameters, i.e., no more than in the Capella example shown in \S \ref{sub:PsdSolution}. This may not be surprising given that the least squares fitting procedure used to extract the diffraction profile employs an analytic approximation (Eqn.\,\ref{eqn:AnalyticApproximation06}) that has 3 free parameters and yields satisfactory fits.

\section{Summary}
\label{sec:Summary} 
\par In summary, we have demonstrated retrieving and analyzing stellar aureole profiles for the purpose of determining the particle size distributions in ice clouds. (1) We have collected a useful stellar aureole database covering different atmospheric conditions and cloud optical depths. (2) Stellar aureole profiles have clearly been detected and measured through a wide range of cloud conditions. (3) The aureole profiles can be followed out to $\sim$$0.2^\circ$ from stars ($\sim$$0.5^\circ$ from Jupiter). (4) The stellar aureoles from cirrus that we have examined have very distinctive profiles, typically being approximately flat out to a critical angle, followed by a steepening power-law decline with a slope less steep than $-3$. (5) The retrieved phase functions cover the range of sizes from $\sim$$50 \,{\rm \mu m}$ to $\sim$$400 \,{\rm \mu m}$. 

\par In the process, we noted that (6) the relation between the diffraction phase functions of complicated ice crystal habits and the Fourier transforms of their projected areas results in a simple wavelength scaling relationship. We compared the phase functions for a variety of crystal habits with the same size metric. We defined the ``area diameter'' and found that (7) the phase functions for different habits are very similar for the same area diameter and (8) could be approximated using a simple analytic formula. We found a similar analytic approximation to represent possibly mutltiply scattered diffraction aureole profiles. (9) We utilized this simple aureole approximation along with a Gaussian function for the PSF and a constant for the background to separate the aureole profile data into these components. (10) We developed a technique for deconvolving multiply scattered diffraction profiles to find the phase functions from which (11) we were able to retrieve reasonable PSDs using constrained numerical inversion. 

\par The next steps are (a) to develop an instrument for the routine, automatic measurement of thin cirrus microphysical properties, (b) to deploy it at instrumented sites, such the Atmospheric Radiation Measurement facilities operated by the Department of Energy, and (c) to compare its cirrus microphysical property retrievals with those of other instruments. In this latter category we include specifically (d) independent validation of the retrieved size distributions of cirrus ice crystals by comparison with coincident {\it in situ} measurements. In the overall process, a number of instrument improvements are anticipated. For example, we expect that the angular dynamic range of the aureole measurements, which is closely related to the range of particle sizes measured, can be increased by a factor of 2 to 3. Several ways of doing this include (i) reducing the background level by locating the instrument to a darker area, (ii) reducing the PSF width by improving the camera optics, and (iii) using multiple color filters to take advantage of the dependence of diffraction on wavelength (Eqn.\,\ref{eqn:WavelengthScaling1}). In connection with instrument development it will be important to investigate the sensitivity of the phase function retrieval to noise in the radiance measurements.

\section{Acknowledgements}

The authors acknowledge and thank Rickey Petty and the Atmospheric System Research Program in the Climate and Environmental Sciences Division of the Department of Energy for support through Phase I SBIR grant DE-SC0006325. The authors also thank the anonymous reviewers for their recommendations, particularly their encouragement to take into account the effects of multiple scattering. Finally, the authors thank Brent Holben and his staff for their work in establishing and maintaining AERONET and especially the site at Goddard Space Flight Center. 

\section{Appendix - Multiple Scattering Deconvolution}
\label{sec:MultipleScatteringDeconvolution}

\par Small-angle, multiple scattering has been studied extensively for well over half a century, particularly in the context of particle or electron scattering from beams (e.g., and references therein, Bethe, 1953). Mathematically the problem involves convolution in the forward direction, or deconvolution in the case of determining the single scattering pattern from measurements (e.g., Misell and Burge, 1969; Jones and Misell, 1967). Assuming cylindrical symmetry in the aureole profiles, we adopt an approach that takes advantage of the Hankel transform (Bracewell, 2000).

\par Consider plane-wave radiation from a distant point source, e.g., a star, of irradiance $S_0 \, ({\rm W / cm^2 / \mu m})$ incident normally on a uniform, plane-parallel, particulate layer, e.g., cirrus. Let $L(\theta) \, ({\rm W / cm^2 / sr / \mu m})$ be the total radiance (luminance) scattered by the particulate layer at scattering angle $\theta$. The total or multiple scattering probability $Q_{ms}(\theta) \, {\rm (sr^{-1})}$ per unit solid angle into an annular ring at scattering angle $\theta$ is:
%
\begin{equation} 
Q_{ms}(\theta) ~= ~\frac{ L(\theta) }{ S_\circ }
\label{eqn:TotalScatteringAureole01}
\end{equation}
\noindent It is convenient to discretize the scattering angle domain, i.e., replacing continuous functions with vectors. Let ${\bf Q_{ms}}[i]$ be the vector element representing $Q_{ms}(\theta_i)$ at scattering angle $\theta_i$, where $1 \le i \le n$ and covers the range from $0 \le \theta \le \theta_{max}$. Then Eqn. \ref{eqn:TotalScatteringAureole01} can be written as:
%
\begin{equation} 
{\bf Q_{ms}} ~= ~\frac{ {\bf L} }{ S_\circ }
\label{eqn:TotalScatteringAureole02}
\end{equation}

\par Consider the equation for single scattering in the forward direction from radiaion incident normally on a uniform, plane-parallel, particulate layer (e.g., pg. 302, Liou 2002). The singly scattered aureole radiance ${\bf L_{ss}} \, ({\rm W/cm^2/sr/\mu m})$ is:
%
\begin{equation} 
{\bf L_{ss}} ~= ~\frac{ \tau ~e^{- \tau } ~S_\circ } {4 \pi } ~{\bf P} ~= ~\tau ~e^{- \tau } ~S_\circ ~{\bf Q} 
\label{eqn:SingleScatteringAureole01}
\end{equation}
\noindent where $\tau$ is the optical depth measured normal to the layer and ${\bf P}$ is the phase function normalized to integrate to $4 \pi$, ${\bf Q}$ is the phase funnction normalized to integrate to $1$, and the single scattering albedo is omitted assuming conservative scattering. Or, dividing ${\bf L_{ss}}$ by $S_\circ$ we find that:
%
\begin{equation} 
{\bf Q} ~= ~\frac{ {\bf L_{ss}} }{ \tau ~e^{-\tau} ~S_\circ } 
\label{eqn:SingleScatteringAureole02}
\end{equation}

\subsection{Poisson Model}
\label{sub:Poisson}

\par Approximate multiple scattering as a Poisson process (e.g., Ning et al. 1995). Then ${\bf Q_{ms}}$ is the sum over the probabilities of all orders of scattering: 
%
\begin{equation} 
{\bf Q_{ms}} ~= 
 \Pi(1,\mu) ~{\bf Q} ~+ 
~\Pi(2,\mu) ~{\bf Q} \otimes {\bf Q} ~+
~\Pi(3,\mu) ~{\bf Q} \otimes {\bf Q} \otimes {\bf Q} + ...
\label{eqn:Poisson01}
\end{equation}
\noindent where $\otimes$ represents the 2-dimensional convolution operator and $\Pi(n,\mu)$ is the Poisson probability for $n$ scatterings:
%
\begin{equation} 
\Pi(n,\mu) ~= ~\frac{ e^{-\mu} ~\mu^n }{ n! } ~= ~\frac{ e^{-\tau} ~\tau^n }{ n! }
\label{eqn:Poisson02}
\end{equation}
\noindent and $\mu$ is the average number of scatterings, which is equal to the optical depth $\tau$. For the particles of interest the forward scattering radiance forming the aureole is much more intense than scattering at other angles. Consequently the integral in the convolution operator $\otimes$ can be approximated by an integral that goes to infinity:
%
\begin{eqnarray} 
f(\theta) ~\otimes ~g(\theta) ~& \approx & ~\int_0^\infty ~\int_0^{2\pi} ~f(\theta) \times \nonumber \\
& & ~g(\theta^2 -\theta'^2 - 2 \theta \theta' \cos \phi' ) ~\theta' ~d\theta' ~d \phi'
\label{eqn:Poisson03}
\end{eqnarray}
\noindent where $\phi$ is an azimuthal angle and the small angle approximation, $\sin{\theta} \approx \theta$, has been employed. 

\par It is convenient to apply the Hankel transform to Eqn.\,\ref{eqn:Poisson01} so that the convolution operation is replaced by multiplication in the transformed domain using the convolution theorem for Hankel transforms. Let $\mathcal{H}\{\}$ be the Hankel transform operator:
%
\begin{equation} 
\mathcal{H}\{f(r)\}(q) ~= ~ 2 \pi ~\int_0^\infty ~f(r) ~J_0(2 \pi q r) ~r ~dr
\label{eqn:Hankel01}
\end{equation}
\noindent where $f(r)$ is a function to be transformed and $J_0$ is the Bessel function of order 0. The inverse Hankel transform is also $\mathcal{H}\{\}$. Let ${\bf \tilde{F}}$ represent the Hankel transform of ${\bf F}$ so that Eqn. \ref{eqn:Poisson01} becomes:
%
\begin{equation} 
{\bf \tilde{Q}_{\rm ms}} ~= 
 \Pi(1,\tau) ~{\bf \tilde{Q}} ~+ 
~\Pi(2,\tau) ~{\bf \tilde{Q}} ~{\bf \tilde{Q}} ~+ 
~\Pi(3,\tau) ~{\bf \tilde{Q}} ~{\bf \tilde{Q}} ~{\bf \tilde{Q}} + ...
\label{eqn:Poisson04}
\end{equation}
\noindent Using Eqn. \ref{eqn:Poisson02}, Eqn. \ref{eqn:Poisson04} becomes:
%
\begin{equation} 
{\bf \tilde{Q}_{\rm ms}} ~= ~ e^{- \tau} ~\left [ ~\frac{ \tau }{ 1! } ~{\bf \tilde{Q}} ~+ ~\frac{ \tau^2 }{ 2! } ~{\bf \tilde{Q}}^2 ~+ ~\frac{ \tau^3 }{ 3! } ~{\bf \tilde{Q}}^3 ~+ ... ~\right ] 
\label{eqn:Poisson05}
\end{equation}
\noindent Eqn. \ref{eqn:Poisson05} can be solved for ${\bf \tilde{Q}}$, which is then transformed to find ${\bf Q}$:
%
\begin{equation} 
{\bf Q} ~= ~\mathcal{H} \{ {\bf \tilde{Q}} \}
\label{eqn:Poisson06}
\end{equation}
\noindent Then using Eqn. \ref{eqn:SingleScatteringAureole01}, ${\bf P}$ is:
%
\begin{equation} 
{\bf P} ~= ~4 \pi ~{\bf Q}
\label{eqn:SingleScatteringAureole03}
\end{equation}

\subsection{First Order Solution}
\label{sub:FirstOrderSolution}

\par Consider Eqn. \ref{eqn:Poisson05} limited to first order scattering:
%
\begin{equation} 
{\bf \tilde{Q}_{ms}} ~\approx 
~ e^{- \tau} \, ~\frac{ \tau }{ 1! } \, ~{\bf \tilde{Q}} 
\label{eqn:FirstOrder01}
\end{equation}
\noindent The solution is simply:
%
\begin{equation} 
~{\bf \tilde{Q}} ~\approx ~\frac{ {\bf \tilde{Q}_{ms}} }{ e^{- \tau} \, \tau }
\label{eqn:FirstOrder02}
\end{equation}
\noindent Use Eqns. \ref{eqn:Poisson06} and \ref{eqn:SingleScatteringAureole03} to find:
%
\begin{equation} 
~{\bf P} ~\approx ~\frac{ 4 \pi \mathcal{H} \{ {\bf \tilde{Q}_{ms}} \} }{ e^{- \tau} ~\tau } ~= ~\frac{ 4 \pi \, ~{\bf Q_{ms}} }{ e^{- \tau} ~\tau }
\label{eqn:FirstOrder03}
\end{equation}

\noindent which, of course, is consistent with Eqns.\,\ref{eqn:TotalScatteringAureole02} and \ref{eqn:SingleScatteringAureole01}, where ${\bf L}$ has been approximated as $\bf{ L_{ss}}$. 

\subsection{Second Order Solution}
\label{sub:SecondOrderSolution}

\par Consider Eqn. \ref{eqn:Poisson04} limited to first and second order scattering:
%
\begin{equation} 
{\bf \tilde{Q}_{ms}} ~\approx 
~\Pi(1,\tau) ~{\bf \tilde{Q}} ~+ 
~\Pi(2,\tau) ~{\bf \tilde{Q}} ~{\bf \tilde{Q}} 
\label{eqn:SecondOrder01}
\end{equation}
\noindent or
%
\begin{equation} 
\Pi(2,\tau) ~{\bf \tilde{Q}}^2 ~+ 
~\Pi(1,\tau) ~{\bf \tilde{Q}} ~- 
~{\bf \tilde{Q}_{\rm ms}} ~\approx ~0
\label{eqn:SecondOrder02}
\end{equation}
\noindent The solution to this quadratic equation is simply:
%
\begin{equation} 
{\bf \tilde{Q}} ~\approx ~\frac{ -\Pi(1,\tau) ~\pm ~\sqrt{ \Pi(1,\tau)^2 ~+ ~4 ~\Pi(2,\tau) ~{\bf \tilde{Q}_{ms}} } }{ 2 ~~\Pi(2,\tau) }
\label{eqn:SecondOrder03}
\end{equation}
\noindent Since the $+$ sign is required, this equation becomes:
%
\begin{equation} 
{\bf \tilde{Q}} ~\approx ~\frac{ \sqrt{ \Pi(1,\tau)^2 ~+ ~4 ~\Pi(2,\tau) ~{\bf \tilde{Q}_{ms}} } ~- ~\Pi(1,\tau) }{ 2 ~~\Pi(2,\tau) }
\label{eqn:SecondOrder04}
\end{equation}
\noindent Substitute $\Pi(n,\tau)$ from Eqn. \ref{eqn:Poisson02} to find:
%
\begin{equation} 
{\bf \tilde{Q}} ~\approx ~\frac{ \sqrt{ e^{-2 \tau} \tau^2 ~+ ~2 e^{- \tau} \tau^2 ~{\bf \tilde{Q}_{ms}} } ~- ~e^{- \tau} \tau  }{ e^{- \tau} \tau^2 }
\label{eqn:SecondOrder05}
\end{equation}
\noindent and simplify:
%
\begin{equation} 
{\bf \tilde{Q}} ~\approx ~\frac{ \sqrt{ 1 ~+ ~2 e^{\tau} ~{\bf \tilde{Q}_{ms}} } ~- ~1 }{ \tau }
\label{eqn:SecondOrder06}
\end{equation}
\noindent Use Eqns. \ref{eqn:Poisson06}, and \ref{eqn:SingleScatteringAureole03} to find:
%
\begin{equation} 
~{\bf P} ~\approx ~\frac{ 4 \pi }{ \tau } ~\mathcal{H} \{ \sqrt{ 1 ~+ ~2 e^{\tau} ~\mathcal{H} \{ {\bf Q_{ms} } \} } ~- ~1 \}
\label{eqn:SecondOrder07}
\end{equation}

\subsection{Series Solution}
\label{sub:SeriesSolution}

\par Consider Eqn. \ref{eqn:Poisson05} again:
%
\begin{equation} 
{\bf \tilde{Q}_{\rm ms}} ~= ~ e^{- \tau} ~\left [ ~1 ~+ ~\frac{ \tau }{ 1! } ~{\bf \tilde{Q}}~+ ~\frac{ \tau^2 }{ 2! } ~{\bf \tilde{Q}}^2 ~+ ~\frac{ \tau^3 }{ 3! } ~{\bf \tilde{Q}}^3 ~+ ... ~- ~1 ~\right ] 
\label{eqn:SeriesSolution01}
\end{equation}
\noindent where we have added and subtracted $1$ within the brackets to make it easier to recognize the series as the expansion of an exponential:
%
\begin{equation} 
{\bf \tilde{Q}_{ms}} ~= ~e^{- \tau} ~\left [ ~e^{ \tau ~{\bf \tilde{Q}} } ~- ~1 ~\right ] 
\label{eqn:SeriesSolution02}
\end{equation}
\noindent Multiply both sides by $\exp( \tau )$ and add $1$:
%
\begin{equation} 
e^{\tau} ~{\bf \tilde{Q}_{ms}} ~+ 1 ~= ~e^{\tau ~{\bf \tilde{Q}}}
\label{eqn:SeriesSolution03}
\end{equation}
\noindent Take the log and solve for ${\bf \tilde{Q}}$:
%
\begin{equation} 
{\bf \tilde{Q}} ~= ~\frac{ 1 }{ \tau } ~\ln ( 1 ~+ ~e^{\tau} ~{\bf \tilde{Q}_{ms}} )
\label{eqn:SeriesSolution04}
\end{equation}
\noindent Use Eqn.\,\ref{eqn:Poisson06} to find:
%
\begin{equation} 
~{\bf Q} ~\approx ~\frac{ 1 }{ \tau } ~\mathcal{H} \{ \ln ( 1 ~+ ~e^{\tau} ~\mathcal{H} \{ {\bf Q_{ms}} \} ) \}
\label{eqn:SeriesSolution05}
\end{equation}
\noindent and Eqn.\,\ref{eqn:SingleScatteringAureole03} to find:
%
\begin{equation} 
~{\bf P} ~\approx ~\frac{ 4 \pi }{ \tau } ~\mathcal{H} \{ \ln ( 1 ~+ ~e^{\tau} ~\mathcal{H} \{ {\bf Q_{ms}} \} ) \}
\label{eqn:SeriesSolution06}
\end{equation}
\noindent Given $L(\theta)$ and $S_0$ and using Eqn.\,\ref{eqn:TotalScatteringAureole02} this equation allows $P(\theta)$ to be calculated. 


\bibliographystyle{apj}
\bibliography{ast4723}

\end{document}